\documentclass{aa}
\usepackage[english]{}
\usepackage{textcomp}
\usepackage{natbib}
\usepackage[utf8]{inputenc}
\usepackage[colorlinks,urlcolor=magenta,citecolor=blue,linktocpage,breaklinks]{hyperref}
\usepackage[varg]{txfonts}
\usepackage{adjustbox} 
\usepackage{xspace}
\usepackage[]{footnote}
\usepackage{booktabs}
\usepackage{orcidlink}
\usepackage{float}
\usepackage{placeins}
\usepackage{txfonts}
\usepackage[dvipsnames]{xcolor}
\usepackage{graphicx}
\usepackage{pgfplots}
\usepackage{tikz}
\usepackage{caption}
\usepackage[colorlinks,urlcolor=magenta,citecolor=blue,linktocpage,breaklinks]{hyperref}
\pgfplotsset{compat=newest}
\usetikzlibrary{backgrounds,arrows,shapes}
\definecolor{purple}{rgb}{0.5,0,0.87}
\begin{document} 
%--------------------------------------------------------------------
   \titlerunning{LIGHTS: At the edge of Source Confusion in ultra-deep imaging surveys}
   \title{LIGHTS. At the edge of Source Confusion with new ultra-deep optical imaging surveys}
   \subtitle{A modern view of the Olbers' paradox}

   \author{S. Zahra Hosseini-Shahisavandi
          \inst{1}\fnmsep\thanks{Corresponding author; 2hs.zahra@gmail.com}
          \orcidlink{0000-0003-3449-2288}
          \and
          Ignacio Trujillo \inst{2}\fnmsep\inst{3} \orcidlink{0000-0001-8647-2874}
          \and
          Giulia Golini \inst{4}
          \orcidlink{0009-0001-2377-272X}
          \and
          Ignacio Ruiz Cejudo \inst{2}\fnmsep\inst{3}
          \orcidlink{0009-0003-6502-7714}
          \and
          Sepideh Eskandarlou \inst{5}
          \orcidlink{0000-0002-6672-1199}
          \and
          Mireia Montes \inst{6}
          \orcidlink{0000-0001-7847-0393}
          \and
          Mauro D’Onofrio \inst{1}
          \orcidlink{0000-0001-6441-9044}
          \and
          Samane Raji \inst{7}\fnmsep\inst{8}
          \orcidlink{0000-0001-9000-5507}
          \and
          Sergio Guerra Arencibia \inst{2}\fnmsep\inst{3}
          \orcidlink{0009-0001-7407-2491}
          \and
          Dennis Zaritsky \inst{9}
          \orcidlink{0000-0002-5177-727X}
          \and
          Ra\'ul Infante-Sainz \inst{5}
          \orcidlink{0000-0002-6220-7133}
          \and
          Nafise Sedighi \inst{2}\fnmsep\inst{3}
          \orcidlink{0009-0001-9574-8585}
          \and
          Zahra Sharbaf \inst{6}
          \orcidlink{0009-0004-5054-5946}
          \and
          Minh Ngoc Le \inst{2}\fnmsep\inst{3}
          \orcidlink{0009-0003-0674-9813}
          \and
          Chen-Yu Chuang \inst{9}
          \orcidlink{0000-0003-2069-9413}
          \and
          Carlos Marrero-de la Rosa \inst{2}\fnmsep\inst{3}
          \orcidlink{0009-0005-1728-8076}
          \and
          Ouldouz Kaboud \inst{6}
          \orcidlink{0009-0007-7712-0683}
          \and
          Nushkia Chamba \inst{10}
          \orcidlink{0000-0002-1598-5995}
          \and
          Richard Donnerstein \inst{9}
          \orcidlink{0000-0001-7618-8212}
          \and
          Garreth Martin \inst{11}
          \orcidlink{0000-0003-2939-8668}
          \and
          Javier Rom\'an \inst{12}
          \orcidlink{0000-0002-3849-3467}
          }
   \institute{Department of Physics and Astronomy, University of Padua,
             Vicolo Osservatorio 3, 35122 Padova, Italy\\
              \email{2hs.zahra@gmail.com}
         \and
             Instituto de Astrof\'isica de Canarias, c/ V\'ia L\'actea s/n,
             38205 La Laguna, Tenerife, Spain
         \and
             Departamento de Astrof\'isica, Universidad de La Laguna,
             38206 La Laguna, Tenerife, Spain
         \and
             School of Physics, University of New South Wales, NSW 2052, Australia
         \and
             Centro de Estudios de F\'isica del Cosmos de Arag\'on (CEFCA), Plaza San Juan, 1, E-44001, Teruel, Spain
         \and
             Institute of Space Sciences (ICE, CSIC), Campus UAB, Carrer de Can Magrans, s/n, 08193 Barcelona, Spain
         \and
             Departamento de F\'isica Te\'orica, At\'omica y \'optica, Universidad de Valladolid, 47011 Valladolid, Spain
         \and
             Laboratory for Disruptive Interdisciplinary Science (LaDIS), Universidad de Valladolid, 47011 Valladolid, Spain
         \and
             Steward Observatory and Department of Astronomy, University of Arizona, 933 N. Cherry Ave., Tucson, AZ 85721, USA
         \and
            NASA Ames Research Center, Space Science and Astrobiology Division M.S. 245-6, Moffett Field, CA 94035, USA
         \and
            School of Physics and Astronomy, University of Nottingham, University Park, Nottingham NG7 2RD, UK
         \and
             Departamento de F\'isica de la Tierra y Astrof\'isica, Universidad Complutense de Madrid, E-28040 Madrid, Spain}
   \date{Received month day, year; accepted month day, year}
% \abstract{}{}{}{}{} 
% 5 {} token are mandatory 
  \abstract{How deep can optical imaging ultimately become before every resolution element contains astronomical sources? This question, closely related to source confusion and ultimately to Olbers' paradox, has become increasingly relevant with the advent of ultra-deep imaging surveys. Using the galaxy and stellar number counts measured in the LIGHTS survey (complemented with Hubble ultra-deep surveys), we estimate the limiting magnitude at which astronomical sources either begin to overlap (assuming a typical ground-based seeing of 1 arcsec) or occupy nearly every image pixel. At the current completeness limit of the deepest optical observations (g, r$\sim$26 mag), the median separation between detected sources is still 3 – 4 arcsec and they cover approximately 30-40\% of the image area. Extrapolating the observed number counts, we predict that ground-based optical images (with seeing around 1 arcsec) reaching a complete magnitude of g, r$\sim$28 mag would become fundamentally limited by source confusion. At present, the integrated light from undetected sources contributes a diffuse background of approximately 30 mag arcsec$^{-2}$. Reaching the confusion limit requires integration times of about 50 h on current 10-m class telescopes, whereas the next generation of Extremely Large Telescopes should achieve it within only a few hours.}
   \maketitle
%-------------------------------------------------------------------
\section{Introduction}

In 1823, Heinrich W. Olbers formulated the apparent contradiction that within the classical cosmological picture of an infinite, static, and eternal universe uniformly populated by stars should produce a night sky as bright as the stellar surfaces themselves \citep{Olbers1823, harrison1964,harrison1987}. In such a universe, the increase in the number of stars with distance exactly compensates for the inverse-square dimming of their individual fluxes, causing every concentric shell of stars to contribute the same amount of light to the sky. The observed darkness of the night sky therefore implies that at least one of the underlying assumptions of the classical cosmological model must fail. Modern cosmology resolves this paradox through the combined effects of the finite age of the Universe and the cosmological redshift associated with cosmic expansion  \citep{hubble1929, wesson1991}.

Although current cosmology has resolved Olbers' paradox in its original form, modern astronomy faces a new version of the paradox. As our telescopes become more powerful and the images we take of the sky become deeper \citep{Beckwith2006, Illingworth2013, Aihara2022,Euclid2025}, the density of sources appearing in our data grows rapidly. For this reason, it is legitimate to ask at what point and at what depth all our optical images of the sky will have no pixels free of light sources. This is the question we seek to answer in this work.

An alternative way to frame the previous question is as follows: At what depth will the optical data reach the source confusion limit \citep{scheuer1957, condon1974}? The answer depends on the most abundant type of source as our images detect fainter and fainter objects. If these faint sources were primarily stars, we could use larger and larger telescopes in space to improve spatial resolution. In theory, there would always be pixels in the image free of light. Thus, the source confusion limit would be more of a technical problem than an astrophysical one. However, we now know that the faintest sources in our catalogs are not point-like (i.e. stellar) sources, but rather increasingly faint galaxies due to their great distance or low intrinsic luminosity \citep[see e.g.][]{1991MNRAS.249..498M,2004PASJ...56.1011K}. In fact, the star-galaxy crossover happens at a magnitude around r$\sim$21-22 mag at high Galactic latitude \citep[see e.g.][]{2001AJ....122.1104Y, Windhorst2011}.

 In this work, using ultra-deep imaging of the sky taken by the LBT Imaging of Galactic Halos and Tidal Structures \citep[LIGHTS; ][]{Trujillo2021A, Zaritsky2024} survey, with a typical spatial resolution of about 1 arcsec from the ground, complemented by very deep data from the Hubble Space Telescope (HST): the  HST Early Release Science \citep[ERS; ][]{Windhorst2011}, the HST Ultraviolet Ultra-Deep Field \citep[UVUDF; ][]{Teplitz2013, Rafelski2015}, and the HST COSMOS field \citep[named as G10;][]{Andrews2017}, we quantify the depth at which source confusion occurs in optical data.  Since the LIGHTS data was collected on the ground without adaptive optics, source confusion in the optical data would be due to the combination of natural seeing conditions and an increasing number of faint sources.

Several methods exist for calculating the depth at which source confusion will occur based on the information currently available in our deepest images. In other works, source confusion is characterized by measuring the number of sources per resolution element to determine if it is high enough to affect the photometry and/or astrometry of the resolved population of objects \citep[see e.g.][]{2001AJ....121.1207H}. As previously mentioned, however, our study employs a much more conservative definition of source confusion. This definition is based on the absence of pixels in the image that are free of light contamination from resolved sources. To measure when source confusion happens, we have different alternatives. One approach is to examine the typical angular separation between sources in the images and extrapolate this behavior to fainter magnitudes. Once this distance is equivalent to approximately twice the Full Width at Half Maximum (FWHM) of the Point Spread Function (PSF), the sources will overlap. Another alternative is to calculate the number of pixels corresponding to the background (image noise) as a function of data depth. This allows us to determine the magnitude of completeness at which there will be no pixels free of signal contaminated by the light from these objects. Both of these approaches are explored in this work.

Analyzing when the source confusion limit occurs in the optical band, whether due to limited ground-based data and/or a limit imposed by the density of natural sources, has implications for the surface brightness limits we can achieve when conducting low-surface-brightness (LSB) science. In this work, we will explore the surface brightness contribution to the background of those sources that potentially exist in our deep images but yet remain undetected.

The paper is structured as follows: Section~\ref{sec:Data} presents the data used in this study. Section~\ref{sec:Building LIGHTS catalogs} describes the methodology used to construct the photometric catalogs. Section~\ref{sec:Estimating the source confusion limit} estimates the source confusion limit in the g and r bands. Section~\ref{sec:Surface brightness contribution to the background due to undetected sources} estimates the contribution to the surface brightness of background sources that remain undetected in our present-day deep optical images. Section~\ref{sec:How much time} explores how much time is necessary for source confusion in the optical bands to be reached by different facilities. Section~\ref{sec:Discussion and conclusions} discusses our findings and summarizes the conclusions. All magnitudes are given in the AB system unless otherwise stated.

%--------------------------------------------------------------------
\section{Data}
\label{sec:Data}
The analysis presented in this paper is based on very deep data from the Large Binocular Telescope (LBT) and the Hubble Space Telescope (HST). We discuss their properties in what follows.

\subsection{LIGHTS survey}

LIGHTS \citep{Trujillo2021A, Zaritsky2024} is an ongoing ultra-deep imaging survey of nearby (D$\sim$15 Mpc) spiral galaxies with stellar masses slightly lower than the Milky Way. The LIGHTS observing and reduction strategy is designed to preserve as much as possible of the LSB features in the images. At the time of writing, the LIGHTS survey comprises a total of 28 nearby galaxies observed in the g and r bands simultaneously with the Large Binocular Cameras  \citep[LBC; ][]{2008A&A...482..349G} mounted on the LBT in Arizona. Currently, around half of the sample has also been observed with the U and z-bands.  For the particular case of this work, we concentrate on the 25 fields which have been fully reduced. The LBC system consists of two cameras, Blue and Red, each composed of 4 CCDs with a pixel scale of 0.224~arcsec/pixel. Each CCD covers approximately $7.8'\times17.6'$, with gaps between the chips of $\sim$18\arcsec. The total effective field-of-view of each LBC camera is around 23$\times$25 arcmin$^2$. The typical amount of time on-source each field was observed is $\sim$1.5h. Typical limiting surface brightness are 31.2 and 30.7 mag/arcsec$^2$ in the g and r bands, respectively ($3\sigma$ in areas equivalent to $10\arcsec \times 10\arcsec$). 

LIGHTS is not a completely homogeneous survey because the depth varies from field to field \citep{Zaritsky2024}. Depth depends primarily on two parameters: the exposure time used to observe the main galaxy in the field and that galaxy's apparent size. The larger the object, the greater the dithering step required to cover it, resulting in slightly lower on-source time per unit area. For point-like objects, \citet{Zaritsky2024} provides the 5$\sigma$ limiting magnitudes in g and r bands using apertures of 2$\times$ the FWHM  for the different fields. The average FWHM of LIGHTS is around 1.2\arcsec\ and the typical 5$\sigma$ limiting magnitudes for point-like sources are 27.3 mag (g-band) and 26.5 mag (r-band).

LIGHTS is significantly deeper than other surveys used to study the structure of nearby galaxies. To illustrate this, in Fig. \ref{fig:compare_lightsandothersurveys}, we show a comparison of the NGC 3972 galaxy and its surrounding environment as observed by LIGHTS and by the Sloan Digital Sky Survey \citep[SDSS DR17;][]{Abdurrofdr17sdss} and by the Dark Energy Camera Legacy Survey  \citep[DECaLS;][]{Dey2019}. In terms of limiting surface brightness, LIGHTS  is expected to surpass \citep{Zaritsky2024} the depth of 10-year coadded images of the Legacy Survey of Space and Time \citep{LSST2009,LSST2019}. However, due to the good seeing quality of the LSST survey, the limiting magnitudes for LSST 10-year coadded images would be similar (26.9 mag; g-band) or a bit better (26.9 mag; r-band) \citep{LSST2009,LSST2019,lsst2022}.

\begin{figure}[htbp]
    \centering
    \includegraphics[width=0.99\linewidth]{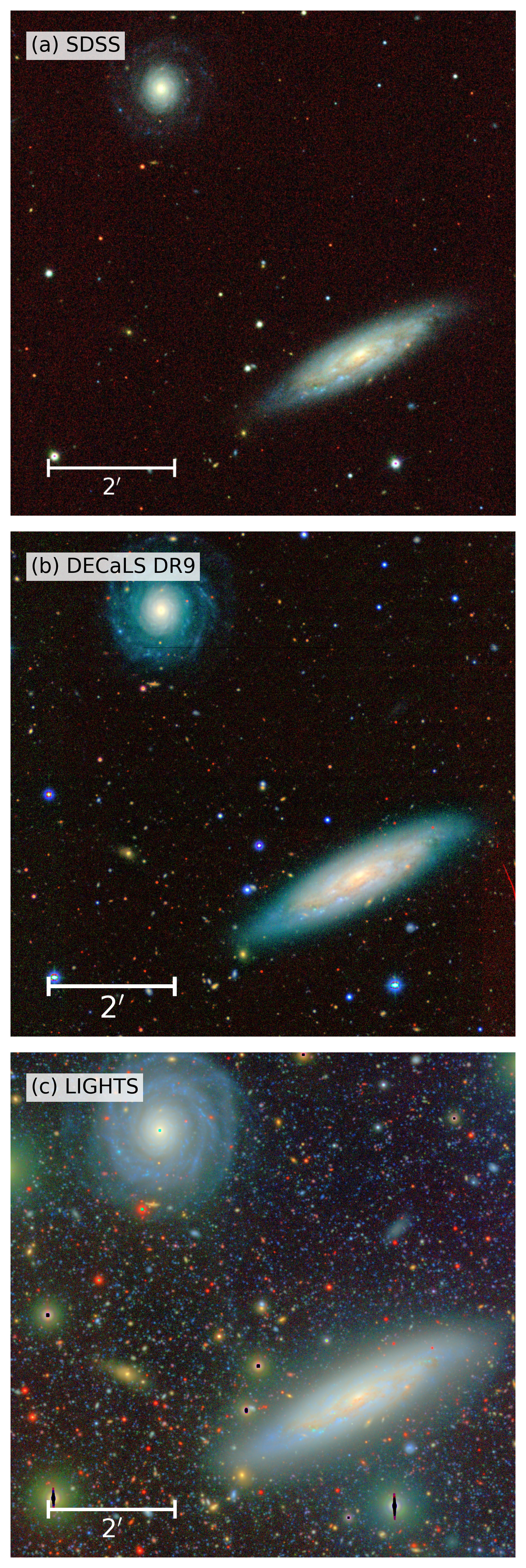}
\end{figure}
\noindent
\begin{minipage}{\linewidth}
    \captionsetup{type=figure}
    \caption{From left to right: The galaxy NGC 3972 and its surrounding region, as observed by the following surveys: SDSS DR17, DECaLS DR9, and LIGHTS. The color images were created using the g, r, and z Sloan bands, using \texttt{astscript-color-faint-gray} script \citep{Infante-Sainz_2024}.}
    \label{fig:compare_lightsandothersurveys}
\end{minipage}

\subsection{Deep HST surveys}

The Hubble Ultra Deep Field \citep[HUDF;][]{Beckwith2006} images and catalogs were retrieved through the MAST High Level Science Products archive\footnote{\url{https://archive.stsci.edu/prepds/udf}} and are used as a complementary dataset to probe our results to fainter magnitudes that are beyond the depth of the LIGHTS data. Additionally, they help us understand the effect of spatial resolution on source confusion estimation. The data was taken with the Advanced Camera for Surveys (ACS). Our reference filter for the comparison with LIGHTS data is F606W (V$_{606}$). The 5$\sigma$ (with the aperture diameter of 0.2\arcsec) limiting magnitude in this filter is 30.2 mag and the size of its FWHM is 0.081\arcsec. The pixel scale of the image is 0.03\arcsec. We refer to this data as ACS HUDF.

Other reanalyses of the HUDF data were conducted by different groups. Among those, we have the one conducted by the Ultraviolet Hubble Ultra Deep Field imaging \citep[UVUDF; ][]{Teplitz2013}. \citet{Rafelski2015} produced a combined catalog of 11-bands. We have used the one corresponding to the F606W. We refer to this data as ACS UVUDF. Additionally, a more ambitious imaging of the HUDF region was conducted by the eXtreme Deep Field \citep[XDF;][]{Illingworth2013}. Incorporating roughly twice the exposure time of the original reduction, the XDF achieves 5$\sigma$ point-source depths of 29.1--30.3 mag in 0.35 arcsec diameter apertures in 9 filters; for F606W it reaches up to 30.3 mag, and for the combined image it even reaches to 31.2 mag. In this work, we used the number counts derived  by \citet{Driver2016} using the catalogs of ACS F606W of XDF dataset. We refer to this data as ACS XDF. The full set of UVUDF, and XDF data products are also available
through MAST\footnote{\url{http://archive.stsci.edu/prepds/uvudf}}\footnote{\url{http://archive.stsci.edu/prepds/xdf}}.

%--------------------------------------------------------------------
\section{Building LIGHTS catalogs}
\label{sec:Building LIGHTS catalogs}
\begin{figure*}
  \centering
  \input{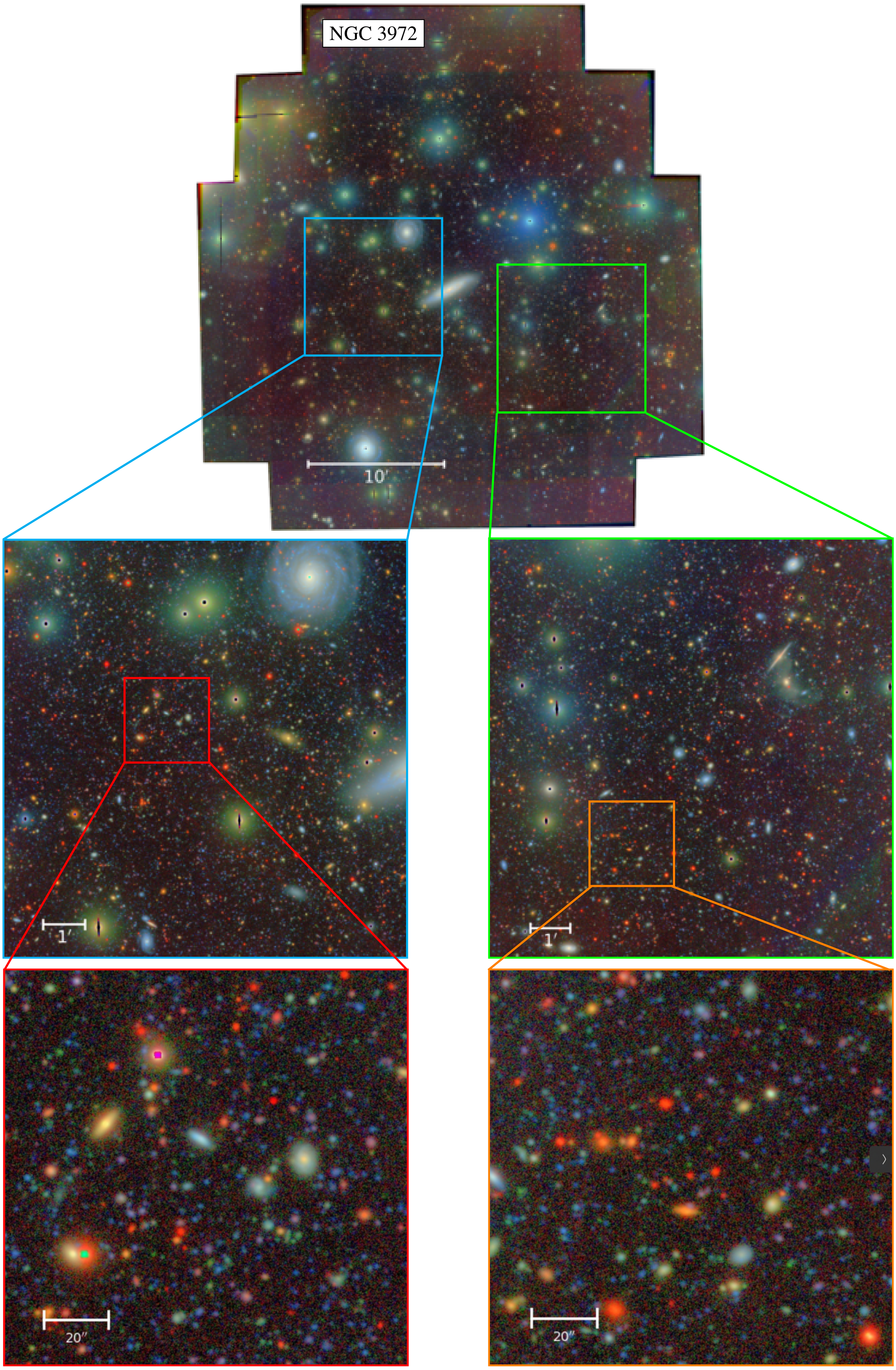}  % Use \input to include the TikZ figure
  \caption{Example of source confusion and crowding in the NGC 3972 LIGHTS field. As shown here, the typical distance between sources is in the order of a few arcseconds. The color image is constructed using the g, r, and z-Sloan bands, using \texttt{astscript-color-faint-gray} script \citep{Infante-Sainz_2024}.}
  \label{fig:lightszoomin}
\end{figure*}

As ground-based optical data gets deeper, the number density of detected sources grows very quickly. To illustrate the current situation in present-day deep optical imaging, Figure  \ref{fig:lightszoomin} shows a number of zoom-ins to specific regions of the LIGHTS field of NGC 3972. As we will quantify later, the typical distance between individual sources is around 3–4\arcsec. 

To estimate the source confusion limit at a given wavelength, the catalog of sources in an image must be as complete as possible. This requires a source extraction code that can accurately characterize the location and magnitude of the sources. In LSB-friendly reduced images, such as those in LIGHTS, building an accurate catalog is particularly complex for a number of reasons. First, the noise of the background is not homogeneous across the entire field of view due to the dithering strategy. Second, the images contain a large number of faint features, ranging from stellar halos of galaxies to extended, very faint background Galactic cirrus. This significantly complicates the identification of pixels that can be considered "truly" background pixels. Another crucial step is cleaning the catalogs that contain sources of artifacts. Fortunately, the way the data is collected in LBT, by simultaneously observing the same sky region in two bands, helps with this cleaning process.

\begin{figure}
  \centering
  \input{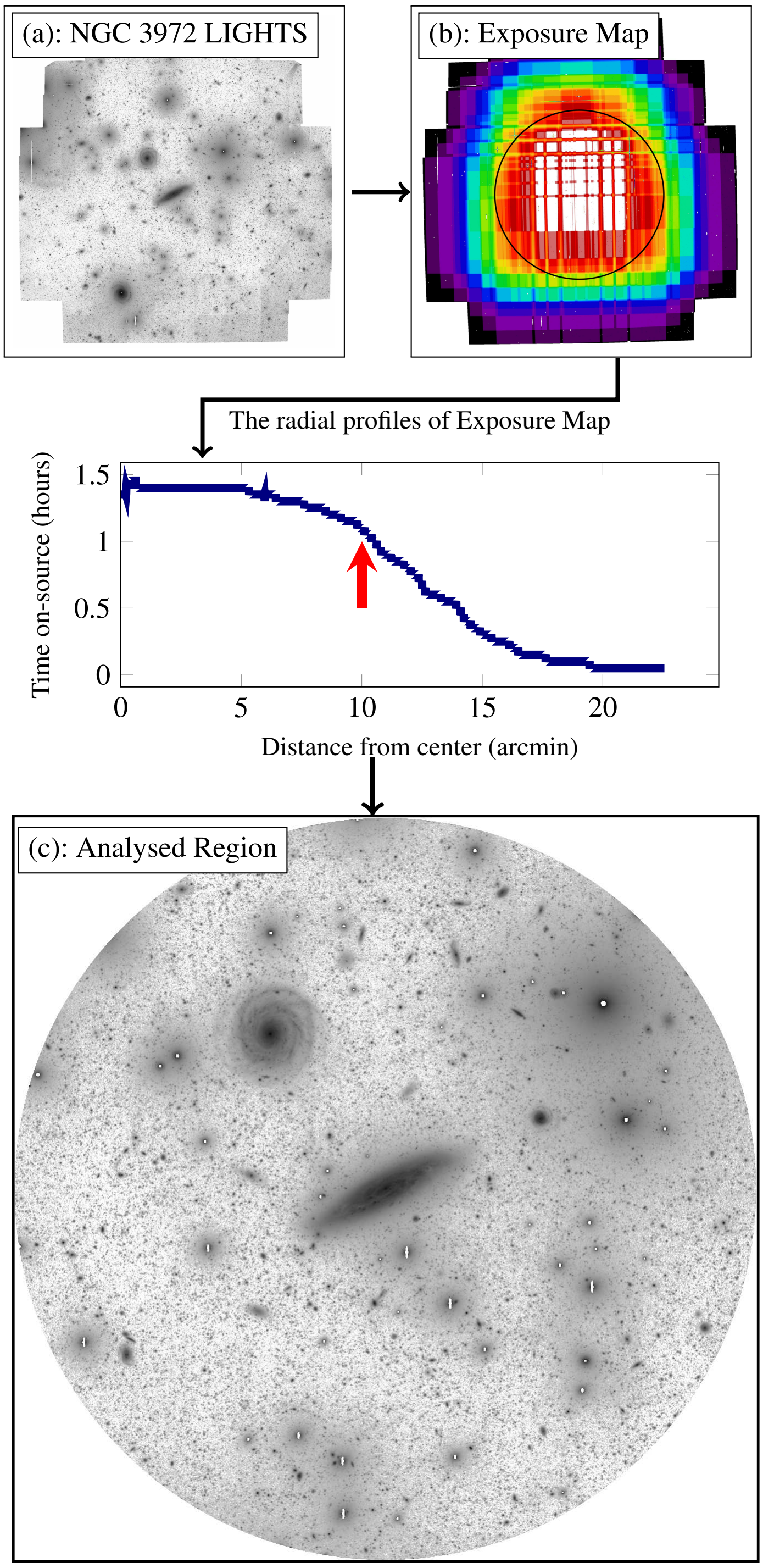}  % Use \input to include the TikZ figure 
  \caption{Selecting the region of the LIGHTS images with 70\% or more on-source time. The radial profile of the exposure map of the final co-added images is used to identify the radial position with at least 70\% on-source time (red arrow). The innermost part of the image (in this plot corresponding to the g-band), as defined by this radius, is used for the source confusion analysis.}
  \label{fig:selection_method}
\end{figure}

To address the inhomogeneity issue in the LIGHTS image across its field of view, we created catalogs of sources only in the innermost region of the combined image for this particular work. To strike a balance between the number of sources we can retrieve from the image and achieving a relatively homogeneous depth in the catalog, we focused on the region of the image that had been exposed for at least 70\% of the observing time. This region is illustrated in Fig. \ref{fig:selection_method}. Using the exposure map of the coadded image, we created an exposure time radial profile. We estimate the radial distance at which the profile reaches 70\% of its value. From then on, we only work with the part of the image that has been exposed for at least that amount of time.

%-----------------------------------------------------------------------------
\subsection{LIGHTS catalogs}
We use the astronomical tools provided by \texttt{Gnuastro}\footnote{\url{https://www.gnu.org/software/Gnuastro}}\ (version 0.23) to build the LIGHTS catalogs. First, we use the \texttt{NoiseChisel} and \texttt{Segment} programs \citep{Akhlaghi2015, Akhlaghi2017, Akhlaghibook2024}. The first code detects signals and separates the background region from the signal. The second code generates a segmentation map. 

In order to extract the LIGHTS catalogs, the images require a number of preparatory steps.   First, the images are affected by saturation at the centers of brighter sources, such as the central galaxy and/or bright stars. These pixels are filled with the maximum value of their neighboring pixels. Second, in order to maximize the signal, the images are convolved with a Gaussian kernel with a FWHM ranging from 2.5 to 4 pixels, depending on the field and filter, and a truncation radius of  5 $\times$ FWHM. A visual inspection was performed to determine the optimal kernel for each field/filter that maximizes the signal. The radius of the kernel is chosen based on the typical extension of the point-like sources in the images, which are the sources we want to maximize in our catalogs.

We take advantage of the two segmentation maps generated by the \texttt{Segment} program: the CLUMPS and OBJECTS extensions; Clumps are associated with local maxima and extend over the neighboring pixels until they reach a local minimum (river), and saved as the CLUMPS extension. The clumps are then grown to a certain threshold over the detections. Based on the strength of the connections (rivers) between the grown clumps, they are considered parts of one object or as separate objects, and saved as the OBJECTS extension \citep{Akhlaghibook2024,Akhlaghi2015}.
The CLUMPS and OBJECTS detections agree for isolated point-like sources against the background of the images. However, when the background under a detected source is generated by the brightness of a larger source, as is the case with a star-forming region in a large galaxy, the CLUMPS detections are identified as part of the larger source. LSB-friendly images, such as those provided by LIGHTS, contain many CLUMPS superimposed on extended, faint sources, such as Galactic cirrus or the extended tails of bright stars. We need to account for these sources in our catalogs. However, we want to avoid including pieces of large galaxies in our catalogs of "true" objects. For example, we want to avoid including CLUMPS detected on top of the central galaxy in an image. Since most central galaxies in LIGHTS are star-forming disks, we count those CLUMPS as potentially star-forming regions and exclude them from our catalogs. For most of our sources (93\%), the \texttt{Segment} identified as CLUMPS coincides with the \texttt{Segment} identified as OBJECTS. Therefore, our catalogs consist of CLUMPS (excluding those found on top of the main galaxy). Since replicability is a critical aspect of this catalog, we have included Table \ref{tab:params} in the Appendix \ref{app:catalogs}. This table lists the values of the parameters in the configuration files for running \texttt{NoiseChisel} and \texttt{Segment} in the present work. The remaining parameters have their default values.

%---------------------------------------------------------------------
\subsection{Removing spurious sources from the LIGHTS catalogs}
\label{sec:Removingspurious}

\begin{figure*}
  \centering 
  \includegraphics[width=0.95\linewidth]{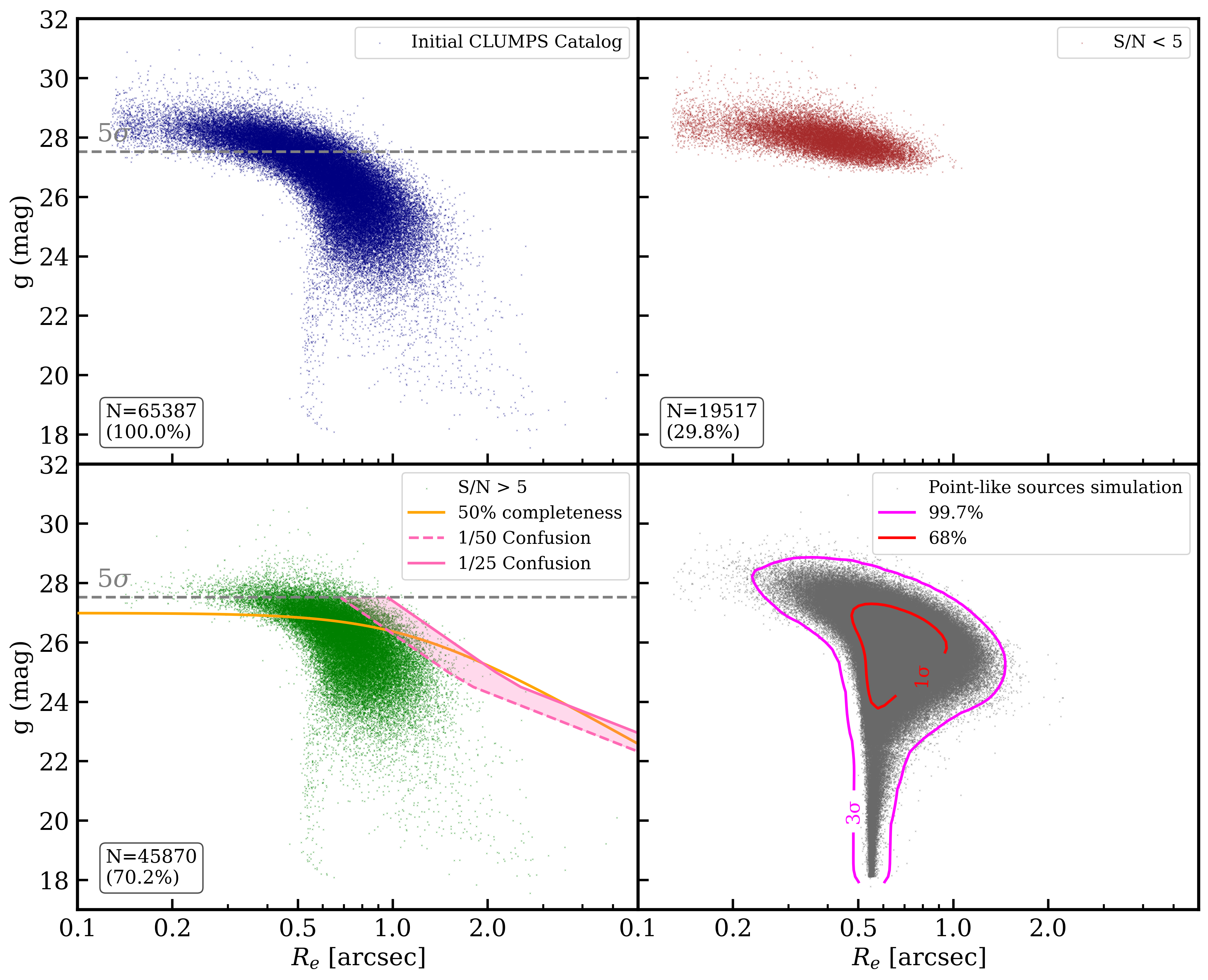}
  \caption{An example of removing the spurious sources from the catalogs. Upper left panel: Effective radius-magnitude plot for all sources detected in the g-band in the LIGHTS NGC 3972 field. The narrow vertical distribution at $R_e$ = 0.55 arcsec and magnitude between 18-22 corresponds to the bright stars in the image. Upper right panel: Same as the previous panel, but showing sources with a signal-to-noise ratio of less than five in our catalogs. Bottom left panel: Final selection; distribution of sources after removing objects with a signal-to-noise ratio of less than five. In this panel, we overplot a number of sensitivity limits that describe the data. The dashed horizontal line corresponds to the 5$\sigma$ limiting magnitude for circular, point-like sources with an aperture of 2$\times$FWHM. The orange line is the 50\% completeness detection limit for exponential-like objects of different magnitudes and effective radii. Finally, the pink region shows when the source confusion limit starts affecting the data, i.e., when more than 2\% of the area is occupied by objects of a given magnitude. See Appendix \ref{app:limits} for details. Bottom right panel: Same plot, but for simulated point-like sources randomly located in the same LIGHTS field. Fainter point-like sources show greater scatter in this plot due to increasing noise when recovering the magnitude and effective radius of these objects. The contours enclose the 68 and 99.7\% of the simulated point-like sources.}
  \label{fig:Simulation}
\end{figure*}

After running \texttt{NoiseChisel} and \texttt{Segment} on the selected regions of the LIGHTS images, we typically identify around 6.5$\times$10$^{4}$ objects per field for the selected region. The upper left panel of Figure \ref{fig:Simulation} illustrates how our methodology works. The left panel uses the LIGHTS NGC 3972 field (g-band) to show the distribution of sources in the effective radius-magnitude plot from our original catalog. This plot displays point-like and extended sources, as well as potential noise detections. As we will show, a significant fraction  ($\sim$30\%) of these detections are spurious and consistent with noise (i.e., they have a signal-to-noise ratio of less than five). To estimate the signal-to-noise ratio of each detection, we consider the number of pixels used to characterize it. We use the measurement \texttt{UPPERLIMIT\_MAG} from the \texttt{Gnuastro} tool \texttt{MakeCatalog}. This measurement provides the level of significance of each detection. This is done by randomly putting the footprint of the object in the regions occupied by the background of the image. The standard deviation ($\sigma$) of the resulting distribution of measured fluxes in those empty regions is used to compute the \texttt{UPPERLIMIT\_MAG} magnitude, which in our case we have chosen to be equivalent to 5$\sigma$.

The upper right panel of Figure \ref{fig:Simulation} shows the locations of all sources in our catalog that are compatible with noise (i.e., those with significance less than 5$\sigma$). As expected, most of these detections are compact and faint. They are consistent with spurious associations of a few bright pixels in the image. After cleaning our catalog of these detections, we are left with a number of sources of high significance, which are distributed in the effective radius magnitude plot, as shown in the bottom left panel of Fig. \ref{fig:Simulation}.

To verify that our removal of spurious sources is effective, we simulated the distribution of point-like sources in our images (see Appendix  \ref{app:simulations}). We generated artificial point-like sources in random locations within the images and determined their effective radii and magnitudes. This allowed us to identify which of our faintest detections were consistent with being artificial sources and which were consistent with being very faint point-like sources. The bottom right plot of Fig. \ref{fig:Simulation} shows the locations of the artificial point-like sources that were randomly distributed in the image. These artificial stars were processed using the same \texttt{NoiseChisel} and \texttt{Segment} configuration as the other sources in the catalogs. Therefore, they are affected by the same crowding and noise problems as the other sources.

The bottom right panel of Fig.  \ref{fig:Simulation} shows the isocontours of the effective radius and magnitude enclosing 68\% and 99.7\% of the artificial star population in the LIGHTS field, NGC 3972. As can be seen, the artificial point-like sources are distributed similarly to the real detections for objects with smaller effective radii (i.e. as the green dots in the bottom left panel of Fig.  \ref{fig:Simulation}). This is reassuring and gives us confidence that most of the removed sources are spurious and not real.

\section{Estimating the source confusion limit}
\label{sec:Estimating the source confusion limit}

We use two different strategies to estimate the depth at which the optical ground-based data would reach its source confusion limit. First, we measure the typical angular distance $\theta$ between sources as a function of their magnitude. As the depth of the images increases, the number of faint sources grows rapidly, and therefore their typical angular distances decrease quickly. By exploring how the typical distance behaves as a function of source magnitude, we can extrapolate the magnitude at which the typical distance reaches a value such that sources overlap. Alternatively, we can measure the source confusion limit by determining the point at which the fraction of pixels occupied by detected sources reaches 100\%. If done correctly, we expect the two estimations to agree on a common limiting magnitude (or imaging depth).

\subsection{Source confusion limit calculation based on the typical angular distance between sources}

To compute the typical angular distance between sources, we proceed as follows: First, we select our catalogs and choose only objects brighter than a given magnitude. Then, for each object, we calculate the angular distance  $\theta$ to its closest neighbor. We perform this calculation using the \texttt{distance-flat} function from \texttt{Gnuastro} that produces a vector with the same number of elements as the number of sources, and represents the angular distance value. To obtain a representative angular distance for the separation between sources in the catalog down to a given magnitude, we calculate the median value of the vector. To avoid potential outliers in the distribution affecting the median (a particular concern when working with bright magnitudes and therefore few sources), we apply a three-sigma clipping cut. Another faster way to find the closest neighbor is to use \texttt{cKDTree} (from \texttt{SciPy}). These two methods produce the same result. 

\begin{figure*}
    \includegraphics[width=1\linewidth]{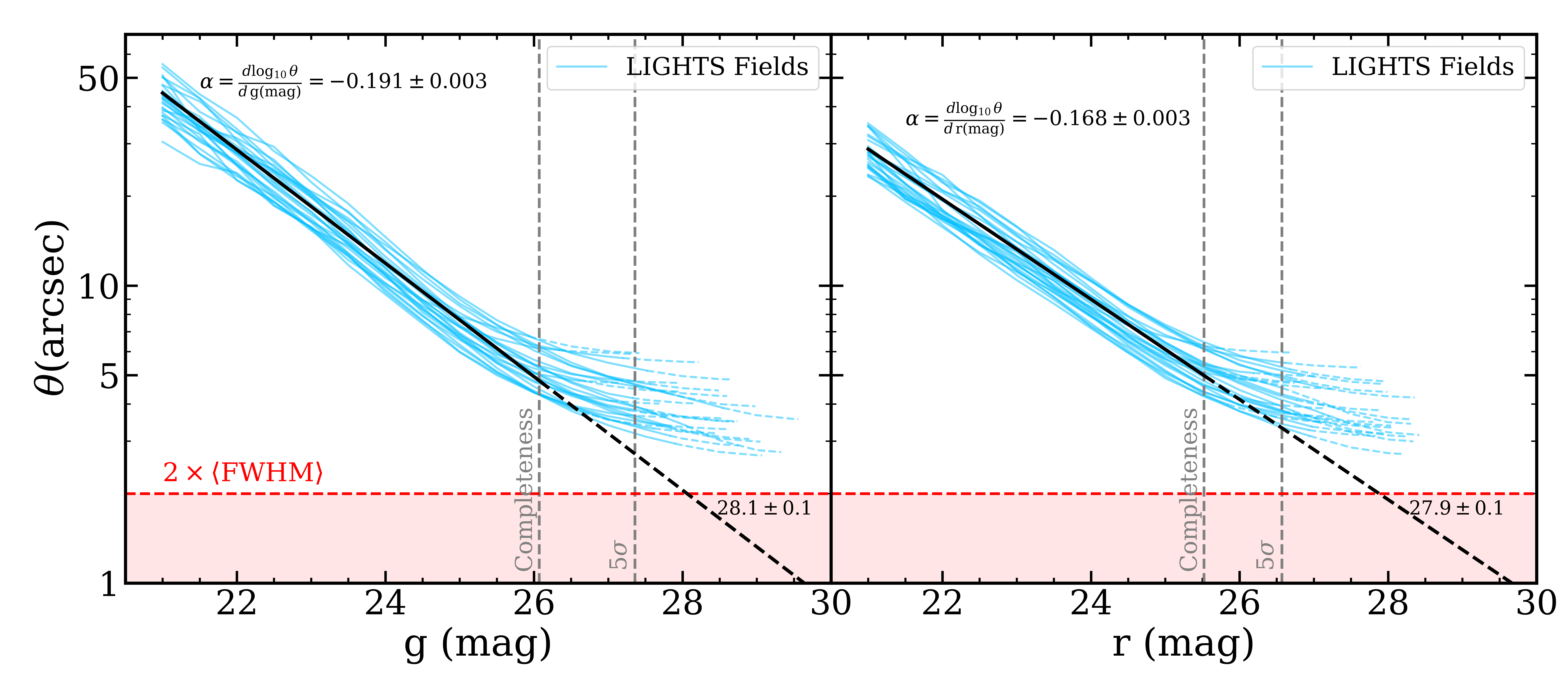}
    \caption{The typical (median) angular distance, $\theta$, between sources for all objects in the LIGHTS catalog brighter than a given magnitude is shown. As objects become fainter, the typical angular separation decreases according to a power-law behavior. Extrapolating this behavior beyond the LIGHTS catalogs' completeness magnitude, source confusion in optical for ground-based images (with seeing FWHM around 1\arcsec) will be reached for data complete down to magnitude 28. The dashed vertical lines show the 5$\sigma$ limiting magnitude for point-like sources and the completeness magnitude, which is more restricted, for both extended and point-like sources in the LIGHTS data.}
    \label{fig:typical_distance}
\end{figure*}

In Fig. \ref{fig:typical_distance} we show the angular distance $\theta$ in the g and r-band for all the sources brighter than a given magnitude, in the 25 LIGHTS fields analyzed in this work. In both bands, $\theta$ steadily decreases with a similar slope ($\alpha_\theta=\frac{dlog\theta}{dm}$) of around $-0.2$. The slope is calculated within the magnitude range where our catalogs are not affected by incompleteness. The completeness magnitudes of very deep surveys such as LIGHTS are not trivial to compute because at the faintest magnitudes the predominant objects are not strictly point-like (i.e. stellar) sources but the catalogs contain a large fraction of faint distant galaxies. For this reason, the 5$\sigma$ limiting magnitude is not representative of the completeness of the catalog, but it is simply an upper limit valid only for strictly point-like sources.

\begin{figure*}
\includegraphics[width=1.\linewidth]{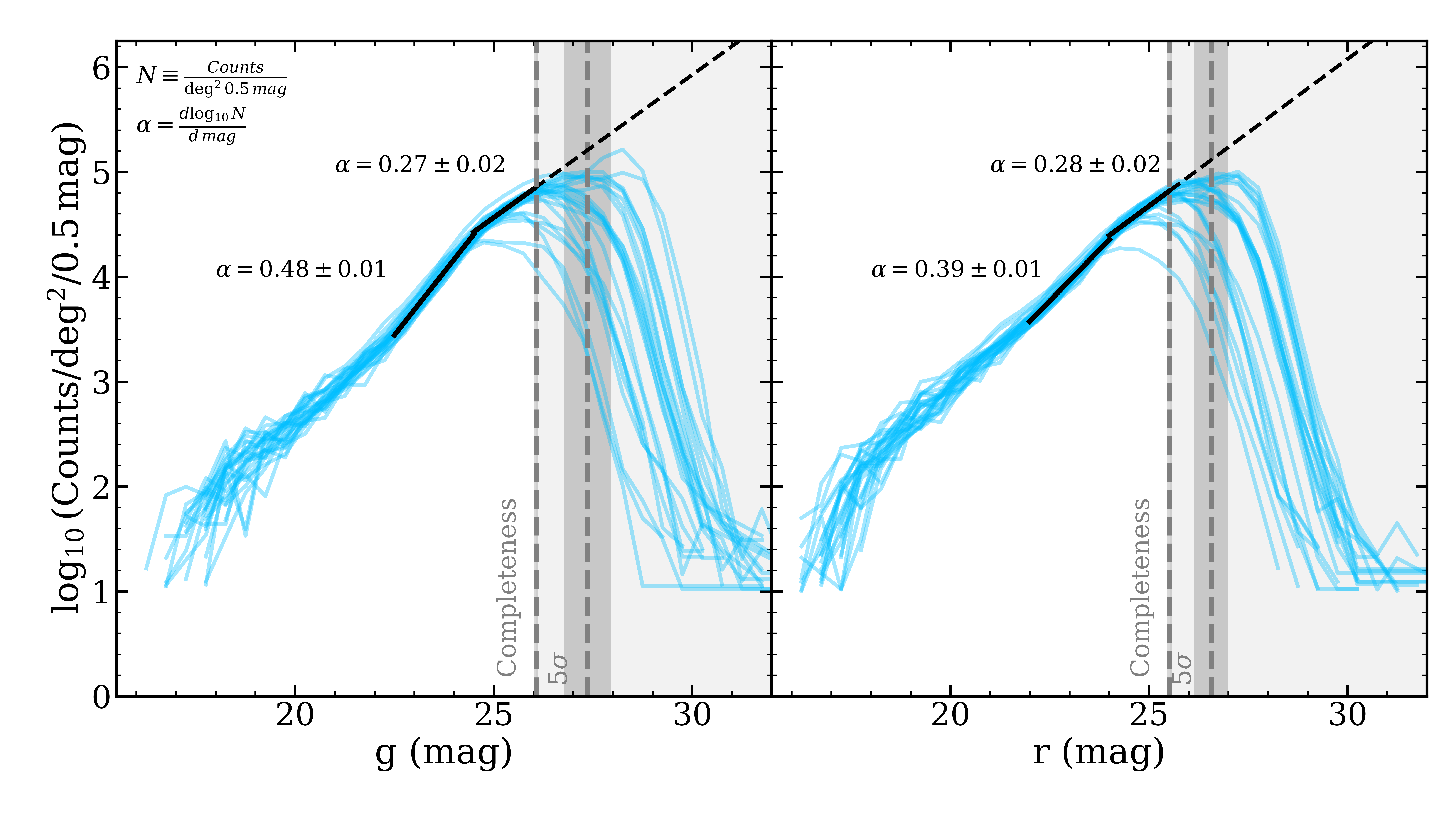}
    \caption{The source number counts for the g- and r-bands in the 25 LIGHTS fields analyzed in this work. The number counts increase rapidly as a function of magnitude. The behavior is similar in both the g- and r-bands. At a magnitude of around 21, the first change in slope occurs, which is associated with the decline in the number of stars in the catalogs. At a magnitude of around 24, another transition in the slope appears, which is likely connected to the change in the slope at the characteristic luminosity L$_{\star}$ of the luminosity function of intermediate redshift galaxies (see the text for details). The extrapolated behavior of the number counts toward fainter magnitudes is shown by the black dashed line. The typical completeness magnitude and the 5$\sigma$ limiting magnitude (for point-like sources) for LIGHTS are shown with dashed vertical lines.}
    \label{fig:Noumbercounts}
\end{figure*}

Due to incompleteness, the behavior of the angular distance between sources in the catalog departs from a power-law decrease and starts to flatten out at a magnitude that depends mostly on the depth of each LIGHTS field.  To obtain a more realistic measurement of completeness magnitude in our catalogs, we examine the number counts of sources as a function of magnitude (see Fig. \ref{fig:Noumbercounts}). As many previous studies have shown \citep[see e.g. ][to name a few]{Windhorst2011,Driver2016,2021MNRAS.503.2033K,2024A&A...691A.240M}, the number counts are well described by a number of power-laws. Around magnitude 21 there is a transition associated with the decrease in the number of faint stars of our own Galaxy \citep[][]{2001AJ....122.1104Y,Windhorst2011}. For magnitudes around 24 (for both g and r), there is another transition \citep{1995MNRAS.273..257M}, changing from a steeper slope: $\frac{dlogN}{dm}$=0.48$\pm$0.01 (22.5$<$g$<$24.5 mag) and $\frac{dlogN}{dm}$=0.39$\pm$0.01 (22$<$r$<$24 mag) to a milder one: $\frac{dlogN}{dm}$=0.27$\pm$0.02 (24.5$<$g$<$26.5 mag) and $\frac{dlogN}{dm}$=0.28$\pm$0.02 (24$<$r$<$26 mag) for fainter sources. These values are in quantitative agreement with those reported in the literature \citep[see e.g.][]{2003A&A...410...17M,2004AJ....127..180C,2004PASJ...56.1011K}.

\begin{figure*}
    \includegraphics[width=0.45\linewidth]{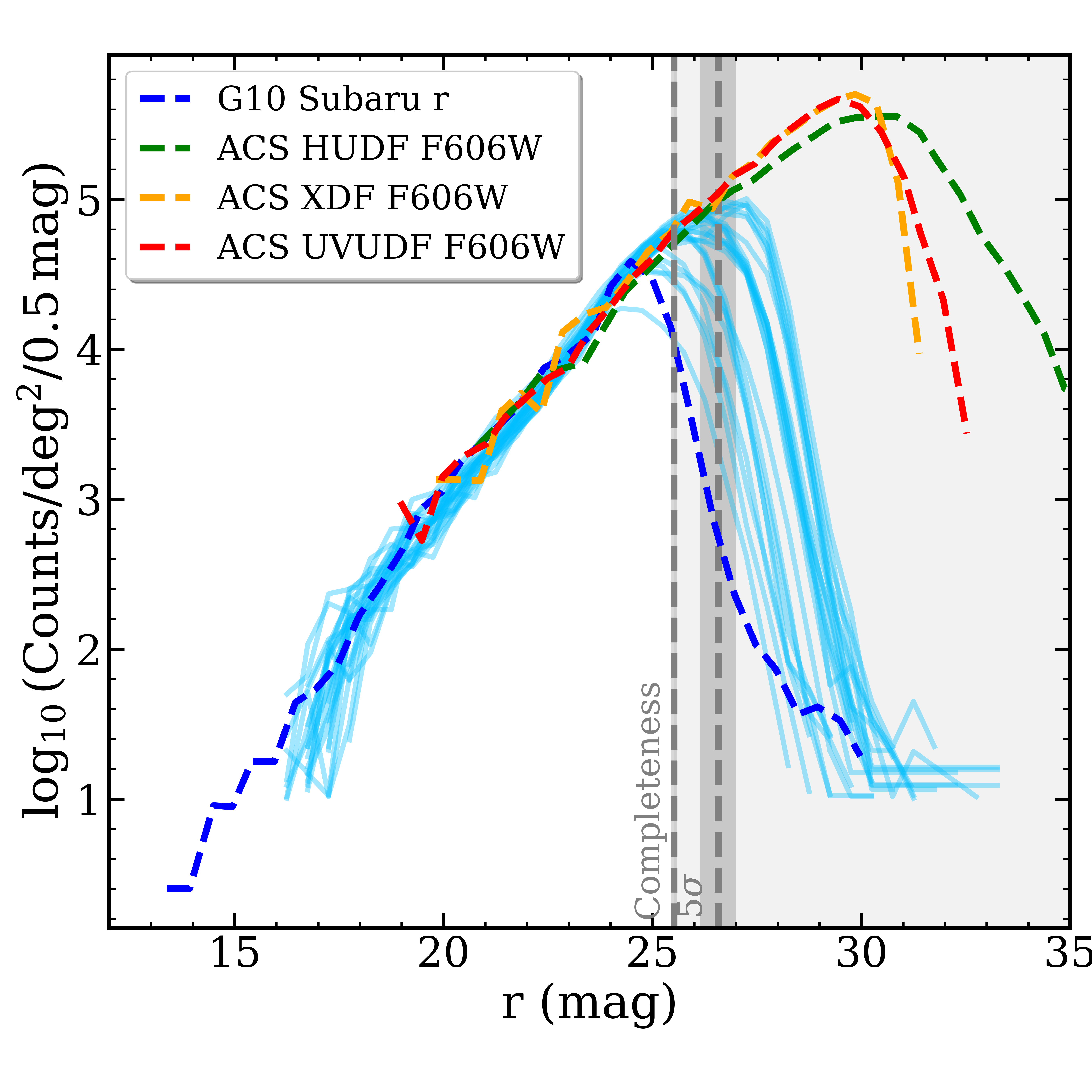}
    \includegraphics[width=0.55\linewidth]{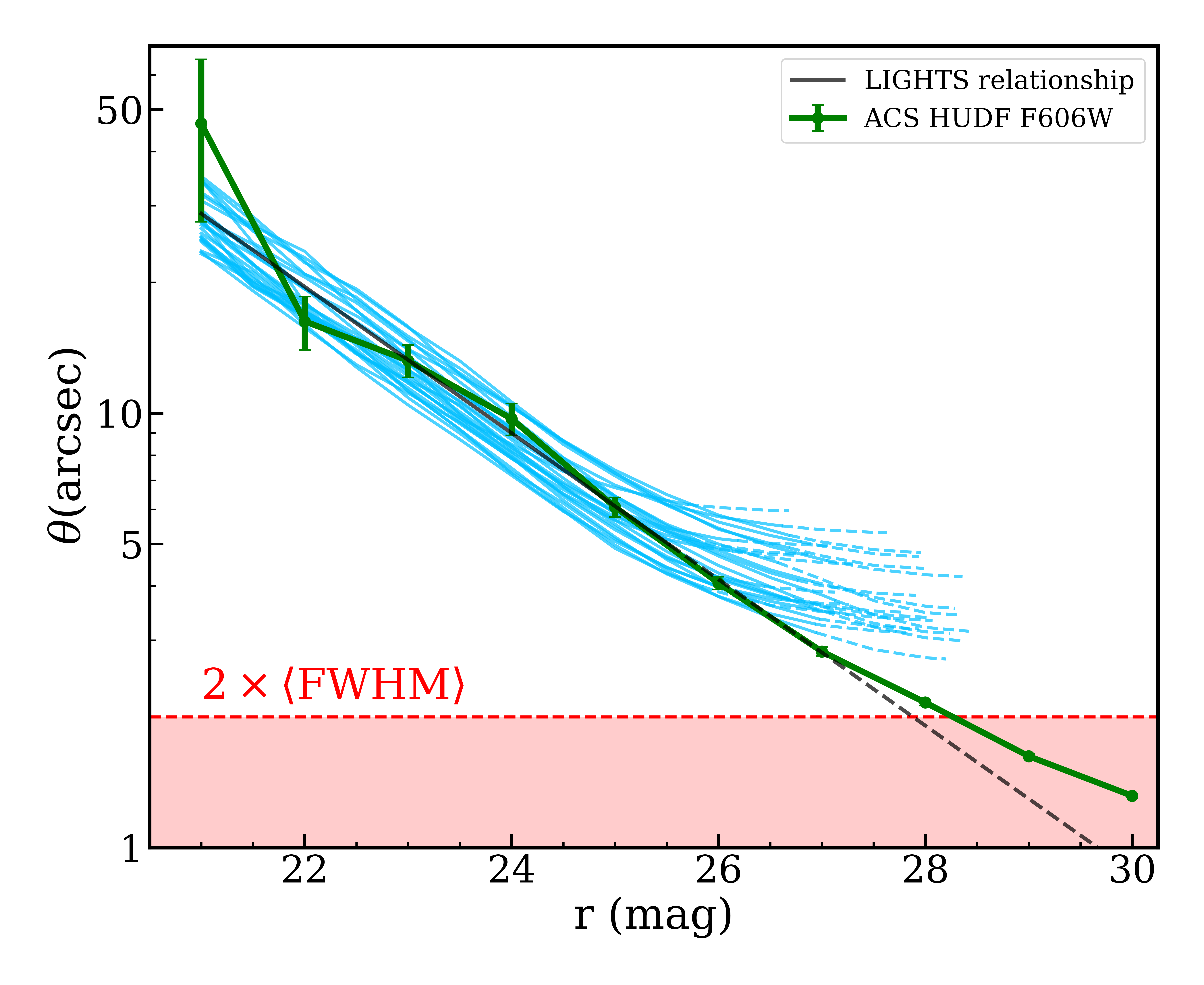}
    \caption{Left panel: Similar to Fig. \ref{fig:Noumbercounts}, but with the addition of number counts from other deep ground-based surveys, such as G10 Subaru \citep[r-band][]{Davies2015, Driver2016,Andrews2017}, and space-based deep fields, such as ACS HUDF \citep[F606W; ][]{Beckwith2006}, ACS UVUDF \citep[F606W; ][]{Rafelski2015} and ACS XDF \citep[F606W; ][]{Illingworth2013}. The LIGHTS number counts are in good agreement with these values. Right panel: Similar to Fig. \ref{fig:typical_distance}, but including the angular separation between sources in deep, space-based catalogs taken by Hubble \citep{Beckwith2006}. The Hubble deeper data follows the step decline depicted by the extrapolation (dashed line) of the LIGHTS data.}
    \label{fig:numbercountwithHST}
\end{figure*}

The origin of the transition around magnitude 24 could have a number of reasons: evolution of the properties of the galaxies at high-z \citep{2001MNRAS.323..795M}, the evolution of the characteristic luminosity L$_{\star}$ of the luminosity function of intermediate redshift galaxies with cosmic time, changes on the geometrical properties of the Universe \citep[see e.g.][]{2000MNRAS.319..807P} and/or the strong effect of the cosmological dimming at high-z \citep{2002MNRAS.329..579C}. Importantly, \citet{2008AdSpR..41.1965W} suggest that at magnitudes fainter than 24, the vast majority of sources are not high-redshift galaxies, but rather, intermediate-redshift, low-luminosity objects. To explore how reliable the slope is in the number counts we find at magnitudes fainter than 24, Fig. \ref{fig:numbercountwithHST} (left panel) overplots the LIGHTS number counts (in the r-band) together with the number counts of sources in the deepest Hubble fields (in the F606W band) \citep{Beckwith2006,Teplitz2013,Illingworth2013}. The agreement with our estimation of the number counts in the common magnitude range is remarkable.

We calculate the completeness magnitude of our catalogs for both point-like and extended sources by comparing LIGHTS's number density counts with those from similarly deep (G10 Subaru, r-band) and deeper photometric catalogs, such as the HUDF, eXtreme Deep Field (XDF) and UVUDF in the F606W filter. The number counts were obtained from \citet{Driver2016}. We use the F606W filter which has a similar response to Sloan r-band.  We define the completeness magnitude of our images as the magnitude at which the LIGHTS counts drop below 90\% of the HUDF counts\footnote{In the case of the g-band, where we cannot make a direct comparison with the HUDF data, we use the deepest LIGHTS field as a reference for calculating the completeness magnitude. The completeness magnitude is 26.1$\pm$0.1 (g-band).}. The typical completeness magnitude in the r-band in the LIGHTS fields is 25.5$\pm$0.1 mag. Now that we have the completeness magnitude of each LIGHTS field, we can fit the relationship between $\theta$ and magnitude within the magnitude range where the catalogs are complete, as shown in Fig. \ref{fig:typical_distance}.

To calculate the magnitude at which we reach the source confusion limit in the optical bands, we extrapolate the power-law behavior of the relationship between $\theta$ and magnitude toward fainter magnitudes (see the dashed lines in Fig. \ref{fig:typical_distance}). We conservatively assume that two sources will overlap if their centers are at a distance of less than two times the FWHM. Therefore, we calculate the source confusion limit when the extrapolated behavior between $\theta$ and magnitude reaches 2$\times$FWHM. For a typical estimation of ground-based data, we assume 1 arcsec of seeing as representative. With this assumption, we find that an image with enough depth to be completed down to magnitude 28 will be limited by source confusion in both the g- and r-bands.
 
A key assumption in our estimation of the source confusion limit is extrapolating the power-law behavior between $\theta$ and magnitude in the magnitude regime where our catalogs are incomplete. To check the reasonability of this assumption, we computed the typical angular distance between sources in the Hubble catalogs. While the Hubble catalogs are noisier at brighter magnitudes due to their limited field of view, we would expect the typical distance in the deep Hubble catalogs to match those derived from LIGHTS at faint magnitudes and to be less affected by incompleteness at fainter magnitudes. The results of this test are shown in Fig. \ref{fig:numbercountwithHST} (right panel). There is very good agreement between LIGHTS and the Hubble catalogs, showing that the power-law extrapolation between $\theta$ and magnitude is a reasonable description at the faintest magnitudes \citep[see also ][]{2021MNRAS.503.2033K,2024A&A...691A.240M}. While the relation between $\theta$ and magnitude using LIGHTS data begins to depart from power-law behavior around magnitude 25.5 (r-band), Hubble's deep data continues the power-law decline down to around magnitude 27.5 (F606W). This is consistent with the fact that the Hubble ultra-deep data are around two to three magnitudes deeper than the LIGHTS data, as seen in the decline in the number counts at the faint end see in Fig. \ref{fig:numbercountwithHST}.

We have also explored the effect of the border of our catalog on the value of  $\theta$ that we derive as a function of the explored magnitude. Since our image has a finite extension, sources located at the border do not "see" sources beyond the edge. Therefore, they estimate the nearest angular distance to objects within the image's border. This implies that the values of $\theta$ measured for such sources are upper limits of the actual values. We have investigated how this "border effect" affects the relationship between  $\theta$ and magnitude. To do so, we maintain the size of our image but calculate the values of  $\theta$ only for objects within 95\% of the radial distance to the border. We keep the original catalog; therefore, those internal sources use all sources to calculate  $\theta$. We do not find any significant change in  $\theta$. In fact, the result remains the same at the faintest magnitude as expected due to the large number of sources used to estimate the typical value of  $\theta$.

Finally, a potential concern regarding our analysis is the representativeness of the LIGHTS fields with respect to the ultimate source confusion caused by mid- and high-z galaxy number counts, as opposed to contamination from faint and distant stars in our own Galaxy. In Appendix Fig.  \ref{fig:typicaldistatlimits_g_r}, we plot the distribution of the analyzed LIGHTS fields in Galactic coordinates. Most of the fields are well above forty-five degrees from the Galactic plane. We did not find any differences in the $\theta$-magnitude plot based on Galactic latitude. Therefore, we conclude that the general trend observed in Fig. \ref{fig:typical_distance} is representative of the source confusion limit based on the background of mid- and high-z galaxies.
 
\subsection{Source confusion limit calculation based on the area occupied by the source detections}

An alternative way to estimate the magnitude of source confusion is to calculate the fraction of pixels occupied by the sources' light. When there are no free pixels left to estimate the background of the image (i.e. the fraction of area with sources reaches unity), the image has reached its source confusion limit. This method complements the one explained above regarding the typical distance of the sources. The first method does not provide information about the extent of the sources, and it treats both point-like and extended sources equally. This assumption oversimplifies the geometry of the sources in the image. Additionally, brighter sources occupy a larger fraction of pixels than fainter ones, regardless of their intrinsic compactness. 

We used the \texttt{CLUMPS} segmentation maps generated by \texttt{Segment} program (\texttt{Gnuastro}) along the catalog to calculate the fraction of pixels in our images occupied by sources. The segmentation program leaves a single pixel line between two adjacent labeled sources. In cases where the sources overlap, these intervening pixels are contaminated by source light. This occurs frequently in the LIGHTS images due to the high source density. Therefore, we fill the borders of empty pixels. We used the \texttt{dilate} function of \texttt{astarithmetic}/\texttt{Gnuastro}. 

\begin{figure*}
    \includegraphics[width=1.\linewidth]{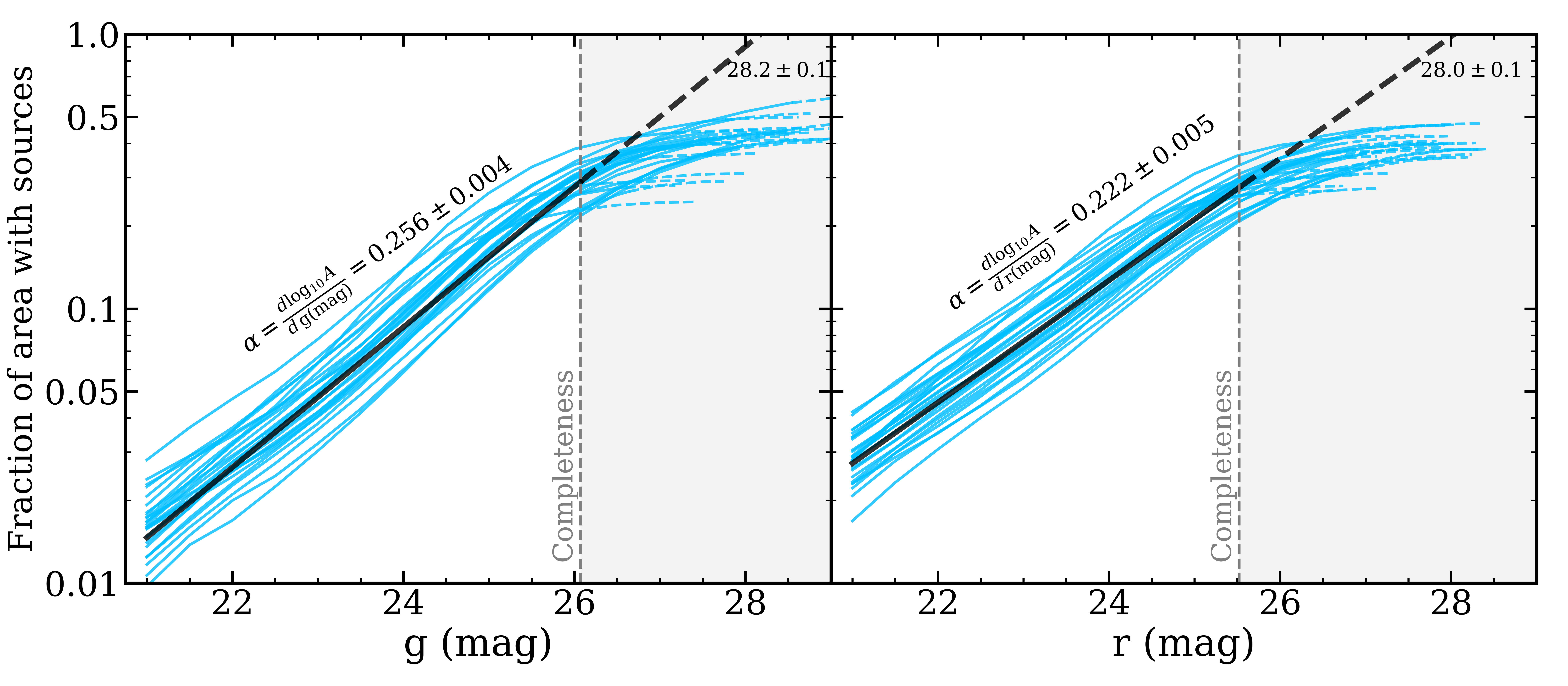}
    \caption{The fraction of pixels in the image occupied by the light of the sources. The occupied area grows according to the depth of the images, exhibiting a power-law behavior. The dashed line extrapolates the power-law regime into the region of magnitudes where the LIGHTS survey is incomplete.}
    \label{fig:NormOccupiedArea}
\end{figure*}

Fig. \ref{fig:NormOccupiedArea} shows the fraction of pixels occupied by sources down to a given magnitude. In both the g and r bands, down to the completeness magnitude of LIGHTS, sources cover between 30 to 40\% of the total pixels in an image. The fraction of pixels with light from individual sources increases as a power law with image depth. We find the following growth:
$\alpha_{A_g} = \frac{d\log A_g}{dg} = 0.256 \pm 0.004$
and
$\alpha_{A_r} = \frac{d\log A_r}{dr} = 0.222 \pm 0.005$,
with $A$ being the fraction of pixels occupied by the light of the sources. Assuming this behavior extends to fainter magnitudes, all pixels in the image would be occupied by light from individual sources if the images were complete down to  28.2$\pm$0.1 mag (g-band) and  28.0$\pm$0.1 mag (r-band). These magnitude values are in very good agreement with the magnitudes for source confusion estimated in the previous method based on the angular distance between sources.

\begin{figure}
\includegraphics[width=1\linewidth]{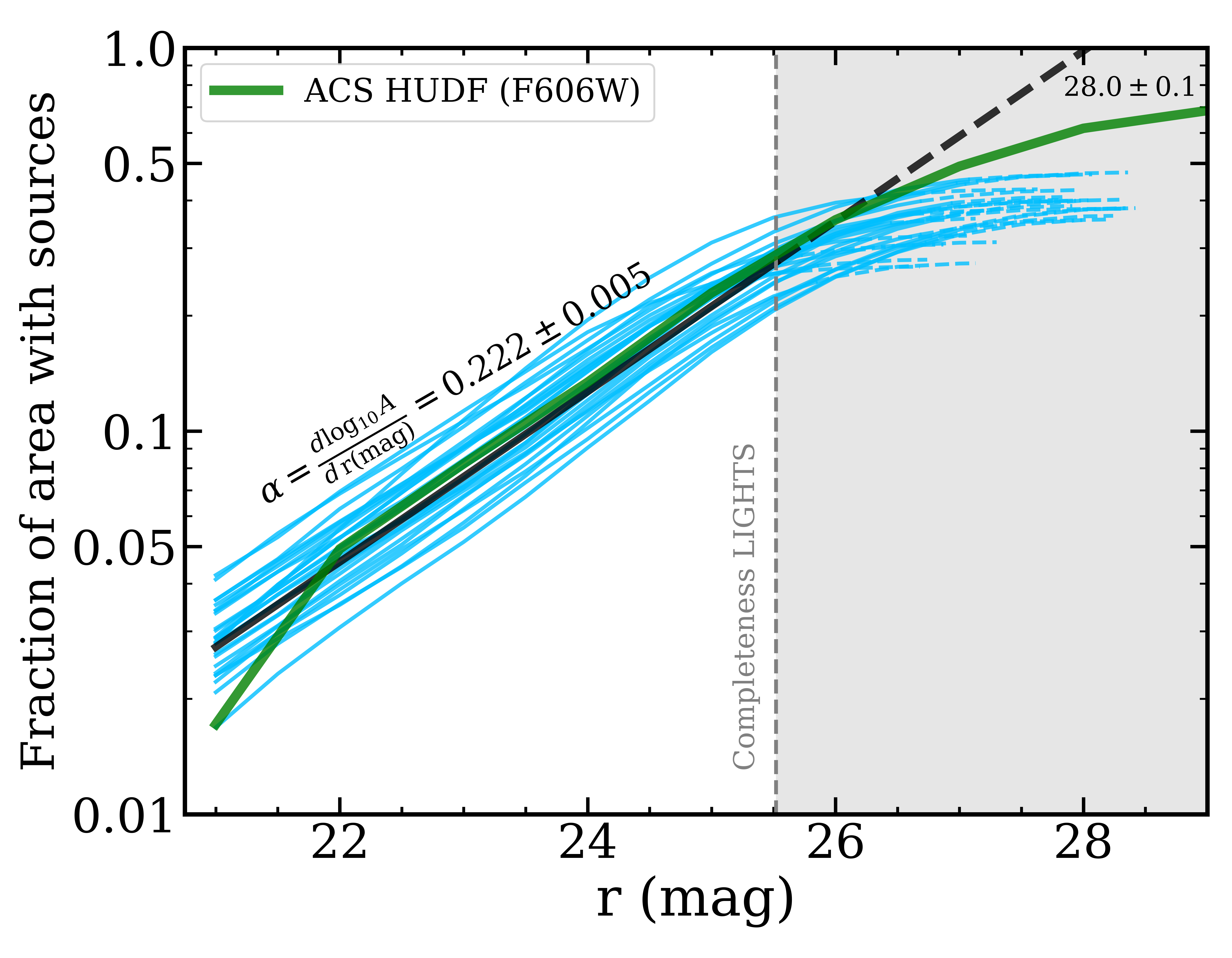}
    \caption{Same as Fig. \ref{fig:NormOccupiedArea} but with the addition of the area occupied by the sources in the Hubble Ultra Deep Field image \citep{Beckwith2006} in the F606W band. The HUDF has been resampled and convolved with the LIGHTS PSF to mimic ground-based conditions (see the text for more information).  The HUDF image used in this study was not processed with a low surface brightness-friendly pipeline. This could explain why the Hubble data departs from the extrapolated behavior of LIGHTS  earlier than expected.}
    \label{fig:WW}
\end{figure}

As we did before with the catalogs of ultra-deep Hubble imaging, we now check whether a power-law extrapolation of the number of pixels occupied by sources in LIGHTS is a good approximation using Hubble data. This time, we use the Hubble Ultra Deep Field image itself \citep{Beckwith2006} in the F606W band. This image has a different pixel and spatial resolution than the LIGHTS data, so it is not representative of the appearance of ground-based data. For this reason, we must transform the HUDF image into an equivalent ground-based image. To accomplish this, we convolve the Hubble data with the LIGHTS PSF \citep[][]{Sedighi2025}. To match the sampling resolution, we change the pixel scale from the ACS camera to that of the LIGHTS data. The details of these steps are shown in Appendix \ref{fig:typical_distance_HUDF}. Once the image is simulated with ground-based conditions from the LIGHTS, we run the same detection algorithm used for LIGHTS images. We apply the same spurious source cleaning procedure that we used with the ground-based data, as discussed in Section~\ref{sec:Removingspurious}. The comparison between the occupied area in the r-band in LIGHTS and the occupied area in the HUDF in the ACS F606W (ground-based simulation) is shown in Fig. \ref{fig:WW}.

Fig. \ref{fig:WW}  shows that, once modeled as a LIGHTS field, the area occupied by the sources in the HUDF follows the extrapolation of the power-law behavior from LIGHTS down to fainter magnitudes. Asymptotically, 70\% of the pixels in the HUDF in the F606W band, as observed from the ground, are occupied by individual sources. There are multiple reasons why 30\% of the HUDF area is not occupied by objects. First, despite the extreme depth of this image, it may be incomplete at faint magnitudes for sources with very low surface brightness. These objects exist in the nearby universe, and their detection is even harder at higher redshifts due to cosmological surface brightness dimming \citep[see e.g.][]{Calvi2014}. Noticeably, the HUDF images have not been reduced using a low surface brightness pipeline, so they are expected to be severely over-subtracted. This implies that the outermost regions of the galaxies in those images are erased, and therefore, the fraction of pixels with light is less than what it should be. An example of this problem is shown in Fig. 18 of \citet{2019A&A...621A.133B}. Finally, the redshifting of sources toward redder wavelengths of higher-redshift galaxies could cause the departure from power-law behavior at fainter magnitudes to be real and not a result of bias, as the first two potential explanations suggest. The quantitative answer to this question is beyond the scope of this paper.  

Finally, in the Appendix \ref{fig:error_SCLM}, we provide a quick estimation of the source confusion limit expected if the images were taken from space-based telescopes. Contrary to what happens on ground-based seeing limited instrumentation, in the space, the ultimate reason for source confusion is the intrinsic apparent size of the high-z sources.

\section{Surface brightness contribution to the background due to undetected sources}
\label{sec:Surface brightness contribution to the background due to undetected sources}

The analysis we are conducting in this work enables us to answer a highly relevant question in the field of low surface brightness science quantitatively. Specifically, we are investigating the contribution of undetected or missed sources in our catalogs to the surface brightness background of our images. This question is important because the flux of these objects contributes to the pixels we use to estimate the "sky background" of our images. The contribution of undetected sources is not the primary source of background in real data because many sources of light contribute to this value. These sources include atmospheric airglow  \citep[$\mu_V\sim$22 mag/arcsec$^2$; ][]{Benn1998}, Zodiacal light \citep[$\mu_V\sim$23 mag/arcsec$^2$; ][]{1998A&AS..127....1L}
, and scattered light from bright sources \citep[$\mu_V\sim$29 mag/arcsec$^2$; ][]{2009PASP..121.1267S,Sedighi2025} all of which are much brighter. However, while airglow and Zodiacal light vary in time and location, they can be averaged out using many individual exposures and removed. Similarly, scattered light is mostly instrumental in origin and can be modeled and subtracted. However, the contribution of undetected sources is different since they are "signal," not "noise." Consequently, these objects always contribute the same amount to the background.  What follows quantifies how relevant this contribution is. 

\begin{figure}
    \includegraphics[width=1\linewidth]{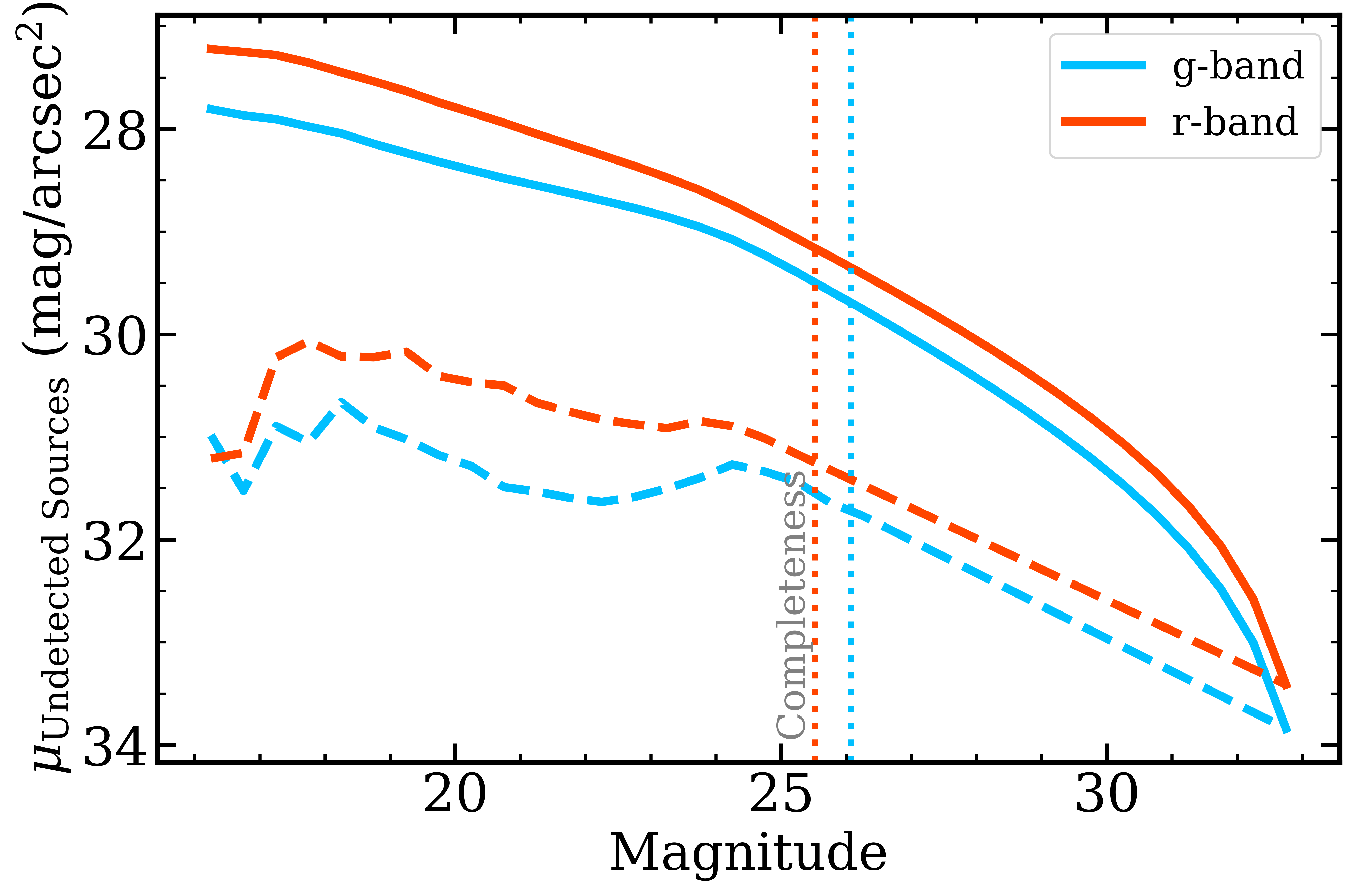}
    \caption{Contribution to the surface brightness of sources of a given magnitude (dashed lines) and the integrated contribution (solid lines) of all the sources fainter than a given magnitude. The vertical dashed lines show the typical completeness magnitudes of the LIGHTS surveys. The contribution to the total background of the sources that remain undetected could be as high as 30 mag/arcsec$^2$.}
    \label{fig:EquivalentSBandTSB}
\end{figure}

We take a simple approach: we estimate the total expected flux per square degree produced by the sources according to the power-law extrapolation of the number counts toward fainter magnitudes. For brighter magnitudes, we use the observed number counts directly by calculating the median value of all LIGHTS number counts. However, when the number of detected sources is below 90\% of the expected number, we use the power-law extrapolation. The results are shown in Fig. \ref{fig:EquivalentSBandTSB}. We present two different estimations of the sources' contribution to the background surface brightness. First, we calculate the contribution of objects of a given magnitude to the total background (dashed lines). Second, we sum the contributions of all objects fainter than a given magnitude to the total background surface brightness (solid lines). The asymptotic values of all the integrated light contribution we find (i.e. the starting point of the solid lines in Fig. \ref{fig:EquivalentSBandTSB}) agree quantitatively with the light contribution of all the galaxies (resolved and non-resolved) to the surface brightness of the sky calculated using deep HST data by \citet{Tompkins2026}. In the g band (F475W), they get 27.6 mag/arcsec$^2$, and in the r band (F625W) they measure 27.3 mag/arcsec$^2$.

One of the most notable findings of our analysis is that the contribution of undetected sources to the background surface brightness in the LIGHTS data is relatively high (i.e., $\sim$30 mag/arcsec$^2$). Therefore, we would expect the surface brightness profiles of the sources in the LIGHTS data (and, more generally, in any deep optical imaging taken from the ground) to exhibit the effects of the unresolved background at these levels of surface brightness. Due to the clumpy nature of this background, we anticipate an increase in noise in the outer regions of resolved galaxies' surface brightness profiles.  This situation could be improved with deeper and/or higher-resolution imaging from space.

 \section{How much time do ground-based telescopes with natural seeing require to reach the source confusion limit?}
\label{sec:How much time}

Deep imaging from the ground taken with large-aperture telescopes has not yet reached the source confusion limit. How far are we from reaching such a limit? To provide a rough estimate of this quantity, we must assume that the number of sources and/or the area occupied by these objects exhibit power-law behavior towards fainter magnitudes, as suggested by the LIGHTS and HUDF results. If so, a deep optical ground-based imaging survey with typical FWHM around 1 arcsec that was complete down to magnitude 28 (in both the g and r bands) would have reached that limit.

\begin{figure}
  \includegraphics[width=\columnwidth]{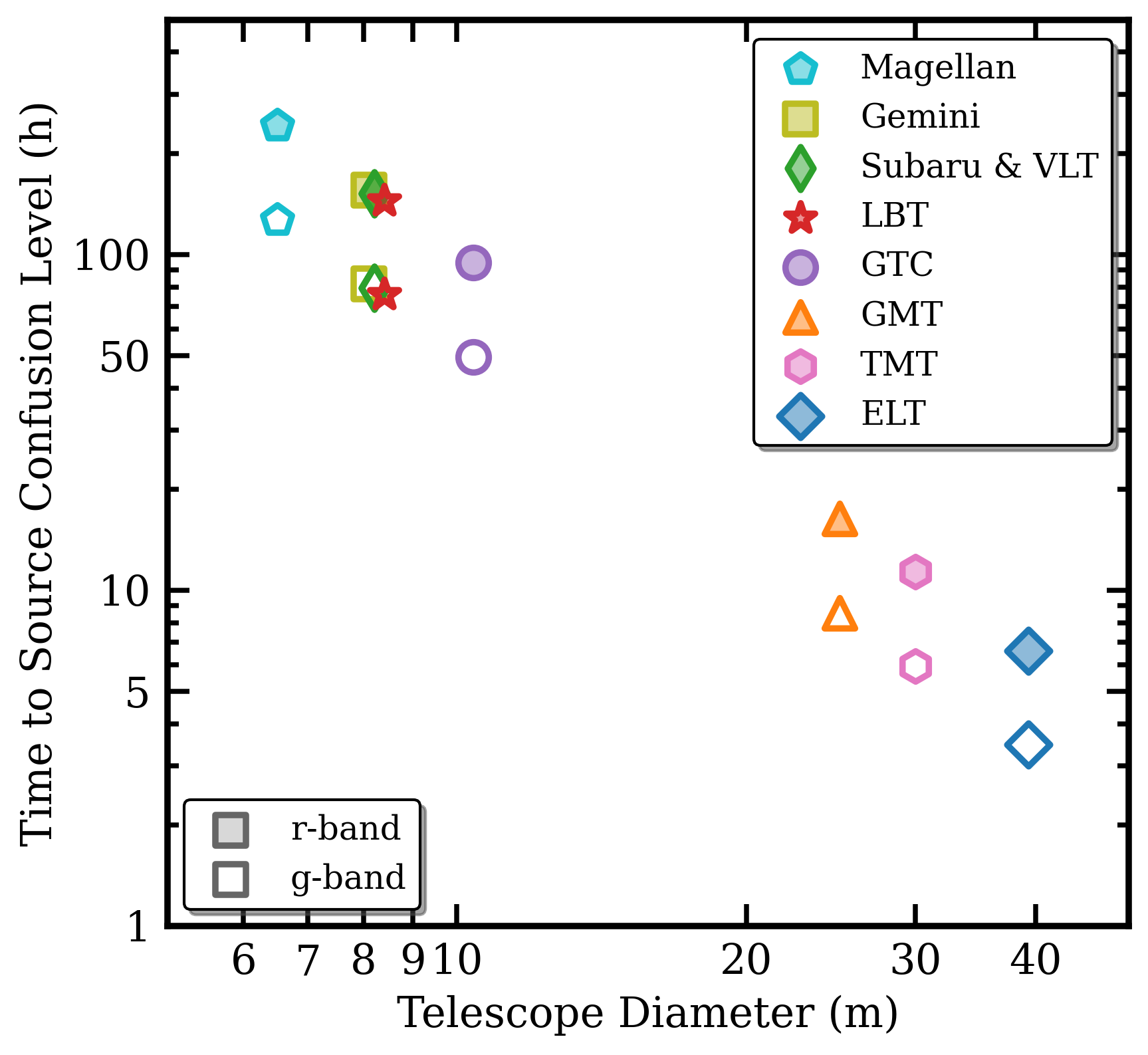}
  \caption{Prediction of maximum exposure time before source confusion level for some ground-based optical telescopes based on the diameter of their mirror.}
  \label{fig:estimate_tmax_before_scl}
\end{figure}

The following calculation is then pretty straightforward. Assuming LIGHTS is representative of typical ground-based seeing conditions (i.e., around 1 arcsec), and knowing the completeness magnitude of our LIGHTS fields and the on-source time allocated to each field, we can estimate the time needed to reach a completeness magnitude of 28, depending on the telescope's aperture. The results of this exercise for a number of current and planned telescopes are shown in Fig. \ref{fig:estimate_tmax_before_scl}. Present-day large optical telescopes, such as the 10.4-meter GTC, would require integrations of around 50 hours (g-band) to reach the source confusion limit with typical seeing of about 1 arcsec. Although this seems like a long time, the next generation of optical telescopes, such as the TMT or ELT, could reach these depths in a few hours.

\section{Discussion and conclusions}
\label{sec:Discussion and conclusions}

Using 25 LIGHTS fields, we explored the extent to which very deep optical ground-based data with 1\arcsec\ FWHM seeing is affected by source confusion. The LIGHTS catalogs are estimated to be complete for both point-like and extended sources down to a magnitude of around 26 in the g and r bands. At these depths, the typical distance between sources ranges from three to four arcseconds, with around 30 to 40\% of the pixels in the images occupied by the light of the sources. The distance between sources and the area they occupy follow a power-law behavior as a function of source magnitude. We explored whether this behavior could reasonably extend to fainter sources by comparing our results with those from the Hubble Ultra Deep Field and found that this hypothesis is reasonable. Therefore, we estimate that future optical images complete down to magnitude 28 will be limited by source confusion. While reaching such depths is currently very time-consuming (requiring around 50 hours or more of observation with our largest 10-meter-class telescopes), this will be achievable in a few hours with the 30- to 40-meter-class telescopes currently under construction.

An important result of this analysis is the contribution of all the undetected, very faint sources to the surface brightness of the background of our present-day, deeper images, well before reaching the source confusion in the optical images. In this paper, we demonstrate that the undetected background of mid- to high-z galaxies has a surface brightness of approximately 30 mag/arcsec$^2$. This background is not perfectly homogeneous, so we expect to see different degrees of clumpiness and color variation from one place to another. In practice, this implies that our low surface brightness extended structures will be noisier than expected.  There are a number of paths we can follow to mitigate this problem. One option is to increase the depth while maintaining the current spatial resolution from the ground, which would reveal fainter sources but would require a significant amount of time. Alternatively, improving the spatial resolution from the ground with adaptive optics or accessing space would improve this situation by around 1 mag/arcsec$^2$. However, this cannot be improved indefinitely by simply making telescopes larger and increasing their spatial resolution. The fainter population of sources in our images is dominated by mid- to high-z galaxies that have an intrinsic extension that barely changes with redshift. This intrinsic extension is the ultimate limit to our source confusion in the optical regime and effectively marks the end of our capabilities to conduct low surface brightness science. 

\begin{acknowledgements}
We would like to thank Professor Rogier Windhorst for carefully reading our manuscript and providing suggestions to improve its quality. We thank Mohammad Akhlaghi for interesting discussions during the development of this work. SZH acknowledges the IAC for hosting her during this research. IT acknowledges support from the State Research Agency (AEI-MCINN) of the Spanish Ministry of Science and Innovation under the grant PID2022-140869NB-I00, financed by the Ministry of Science and Innovation, through the State Budget and by the Canary Islands Department of Economy, Knowledge, and Employment, through the Regional Budget of the Autonomous Community. This research also acknowledges support from the European Union through the following grants: "UNDARK" and "Excellence in Galaxies - Twinning the IAC" of the EU Horizon Europe Widening Actions programmes (project numbers 101159929 and 101158446). Funding for this work/research was provided by the European Union (MSCA EDUCADO, GA 101119830). This work is part of grant CEX2025-001609-S, awarded to the Instituto de Astrofísica de Canarias under the Severo Ochoa Centre of Excellence program and funded by MICIU/AEI/10.13039/501100011033. MM acknowledges support from grant RYC2022-036949-I financed by the MICIU/AEI/10.13039/501100011033 and by ESF+, and program Unidad de Excelencia Mar\'{i}a de Maeztu CEX2020-001058-M, financed by MCIN/AEI/10.13039/501100011033, and by the MaX-CSIC Excellence Award MaX4-SOMMA-ICE. SE acknowledges grant PID2024-162229NB-I00 funded by MICIU/AEI/10.13039/501100011033 and by “ERDF A way of making Europe” (ERDF/EU), by the European Union. RIS acknowledges financial support from the Spanish Ministry of Science and Innovation (MCIN/AEI/10.13039/501100011033 y FEDER, Una manera de hacer Europa) with grant PID2022-138896NA-C54. CMR acknowledges financial support from the Spanish Ministry of Science, Innovation, and Universities (MICIU) through PID2025-174546NB-I00 (TRAMWAYS project).
\end{acknowledgements}
%--------------------------------------------------------------------
\bibliographystyle{aa}
\bibliography{bibliography.bib}
%--------------------------------------------------------------------
\onecolumn
\appendix

\section{Configuration parameters of \texttt{NoiseChisel} and \texttt{Segment}}
\label{app:catalogs}

The following table lists the configuration parameter values used to extract the catalogs in our LIGHTS images with \texttt{NoiseChisel} and \texttt{Segment}. 

%% Table for parameters of the catalog and their definition 
% ----------- Table 1 -----------
\begin{table}[htbp]
 \centering
 \caption{Input parameters of NoiseChisel and Segment}
 \label{tab:params}
 \begin{tabular}{l|c}
  \hline\hline
  NoiseChisel's Parameter  & Input Value\\
  \hline\hline
  tilesize & 20,20\\
  meanmedqdiff & 0.005\\
  interpnumngb & 50\\
  qthresh & 0.25\\
  erode & 1\\
  minskyfrac & 0.5\\
  snquant & 0.95\\
  detgrowquant & 0.85\\
  detgrowmaxholes & 10000\\
  \hline\hline
  Segment's Parameter & Input Values\\
  \hline\hline
  tilesize & 20,20\\
  interpnumngb & 30\\
  objbordersn  & 2.25\\
  clumpsnthresh & 0.48\\
  \hline
 \end{tabular}
\end{table}

% ----------- Table 2 -----------
%\begin{table}[htbp]
% \centering
% \caption{Definition of parameters in catalog}
% \label{tab:definitions}
% \begin{tabular}{l|l}
%  \hline\hline
%  Parameter Name & Definition of Parameter\\
%  \hline\hline
%  RA & Flux-weighted center right ascension.\\
%  Dec & Flux-weighted center declination.\\
%  SN & Signal-to-noise ratio of objects or clumps.\\
%  Half-max-radius & Radius at half the maximum (non-parametric).\\
%  Magnitude & Total magnitude of objects or clumps.\\
%  Other params & Definition of other parameters.\\
%  \hline
% \end{tabular}
%\end{table}

\section{Completeness simulations and source confusion estimations}
\label{app:limits}

The bottom left panel of Fig. \ref{fig:Simulation} shows different lines indicating the limits of our LIGHTS data. The horizontal dashed line corresponds to the 5$\sigma$ limiting magnitude for circular objects with an aperture twice the FWHM of the LIGHTS field, as explained in Section 2.1. The orange line indicates the 50\% completeness limit for objects with an exponential shape and an axis ratio of 0.5. This ratio was chosen as the median axis ratio of randomly inclined disk galaxies. We estimated this line by injecting artificial exponential objects of different magnitudes and effective radii into our LIGHTS images. We followed a procedure similar to those explained in \cite{2006ApJ...650...18T} and \cite{2025PASP..137i4501G}.

\begin{figure}[H]
    \centering
   \includegraphics[width=0.6\linewidth]{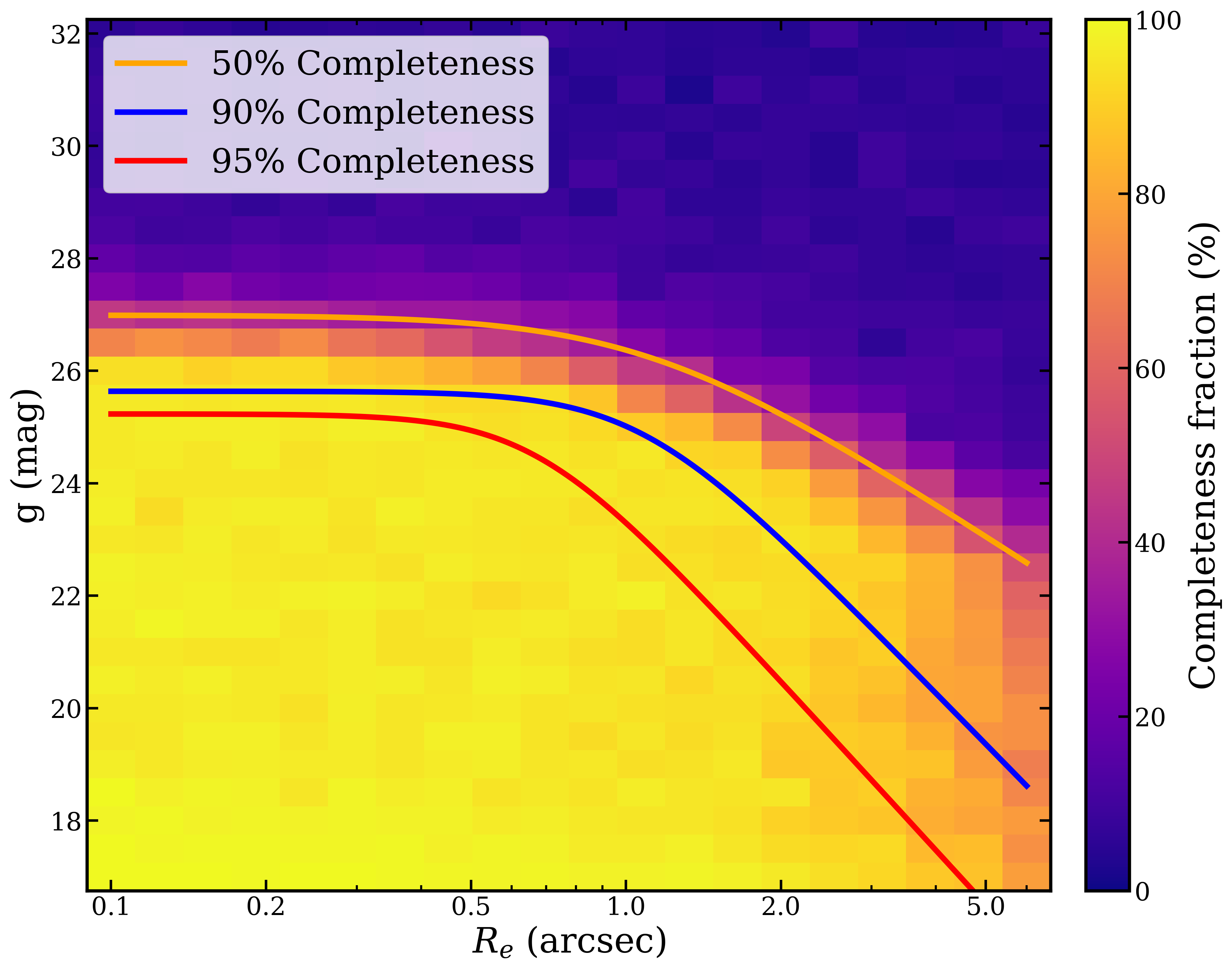}\\
\caption{Plot showing the completeness limits for an object with an exponential profile, an axis ratio of 0.5, and different magnitudes and effective radii. The lines represent different completeness levels ranging from 50 to 95\%. The lines are described by Eq. (1) in \citet{2025PASP..137i4501G}. Each box shows the percentage of artificial objects detected in a given LIGHTS image, depending on their magnitude and effective radius.}
    \label{fig:completeness}
\end{figure}

We used the \texttt{MakeProfile/Gnuastro} program to generate these artificial sources. We added Poissonian noise to the objects. They were also convolved with the PSF of the images before being injected into random locations within the images. We created three hundred artificial galaxies per magnitude and effective radius bin. To avoid overcrowding the data, not all the artificial sources were injected at the same time. In fact, we injected only 620 sources at a time and repeated the process 300 times. Figure \ref{fig:completeness} shows the fraction of artificial objects detected per magnitude and effective radius bin. We use the same tools to detect these artificial galaxies as we do to detect real sources in our images. Some artificial objects are not found due to their very low surface brightness level. Others that could potentially be detected due to sufficient signal-to-noise can be missed due to overlap with real sources. In Fig. \ref{fig:completeness}, we overplot three lines indicating different completeness levels ranging from 50 to 95\%. These lines correspond to the fit of the map using the analytical form provided by \citet[][ their Eq (1)]{2025PASP..137i4501G}.

The source confusion limit of an image is traditionally defined as the flux limit at which sources occupy two percent or more of the image area. Depending on the resolution of the images, two regimes can be distinguished: one when the size of the objects is smaller than the spatial resolution (beam) of the images, and one when the size of the objects is larger. In the first regime, source confusion is called instrumental, while in the second, it is called natural.  Due to the large magnitude range of our LIGHTS objects, our sample expands across the two regimes. At fainter magnitudes, the size of our objects is not representative of their "true" size since they are smaller than the seeing in our data. Conversely, at brighter magnitudes, our galaxies are larger than the seeing. In this paper, we follow the approach of  \cite{2008AdSpR..41.1965W} and \cite{2022ApJ...940L..15K} to estimate the location of the source confusion limits for objects of a given magnitude on the magnitude vs. effective radius diagram (see Fig. \ref{fig:Simulation}). This region is calculated as follows: We estimate the confusion radius R$_{conf}$ at a given magnitude "m" such that the product of the source number density N(m) and the area defined by a source with a radius of R$_{conf}$ is equal to 1/25 and 1/50 (i.e., N(m)$\times$$\pi$R$_{conf,1}^2$=1/25 and N(m)$\times$$\pi$R$_{conf,2}^2$=1/50).

In Figure \ref{fig:Simulation}, the region where source confusion (at a given magnitude) becomes relevant has a significant change in slope around g$\sim$24.5 mag, which coincides with a sharp increase in the number density of sources at that magnitude. At bright magnitudes, the objects are systematically larger, but their number density is so low that they occupy a significantly smaller area of the image, less than 2\%. At faint magnitudes, however, the number density of sources is very high, and their extensions are dominated by the spatial resolution of ground-based data. This means that some galaxies at faint magnitudes start to overlap with other sources in the images, so some could be missed by the detection algorithm.

\section{Artificial point-like source simulations}
\label{app:simulations}

The artificial point-like sources presented in Sec. \ref{sec:Removingspurious} are simulated as follows: For each field and band, we select a bright, non-saturated $\sim$20 mag (G-band) star. We select the representative star of each field by eye, minimizing the number of contaminant sources in its vicinity. We confirm that the selected object is a pure star by checking that its parallax has been measured by Gaia \citep{GaiaDR3}. Once the star has been selected, we crop the star's image using a 3.2 arcsec radius circle (for the representative field in here), almost the radius at which its surface brightness radial profile reaches the surface brightness limit of that field. This ensures that we select an image of the star in which noise does not significantly affect its light distribution. Finally, we normalize the total flux within the cropped image to one. We use this image to simulate stars with different fluxes by scaling it to the desired flux level. In particular, we create artificial stars within the magnitude range  18 to 30 for the g-band and r-band. These stars are randomly located within the area of the LIGHTS images we are investigating. To avoid modifying the number density of the real images and affecting the properties (magnitude and effective radius) retrieved from the simulated stars, the number of simulated stars placed on a real image simultaneously is selected based on their magnitudes. There are very few when the simulated star is bright, and more when it is faint. 

\section{Location of the 25 LIGHTS fields}

\begin{figure}[H]
    \centering
   \includegraphics[width=0.6\linewidth]{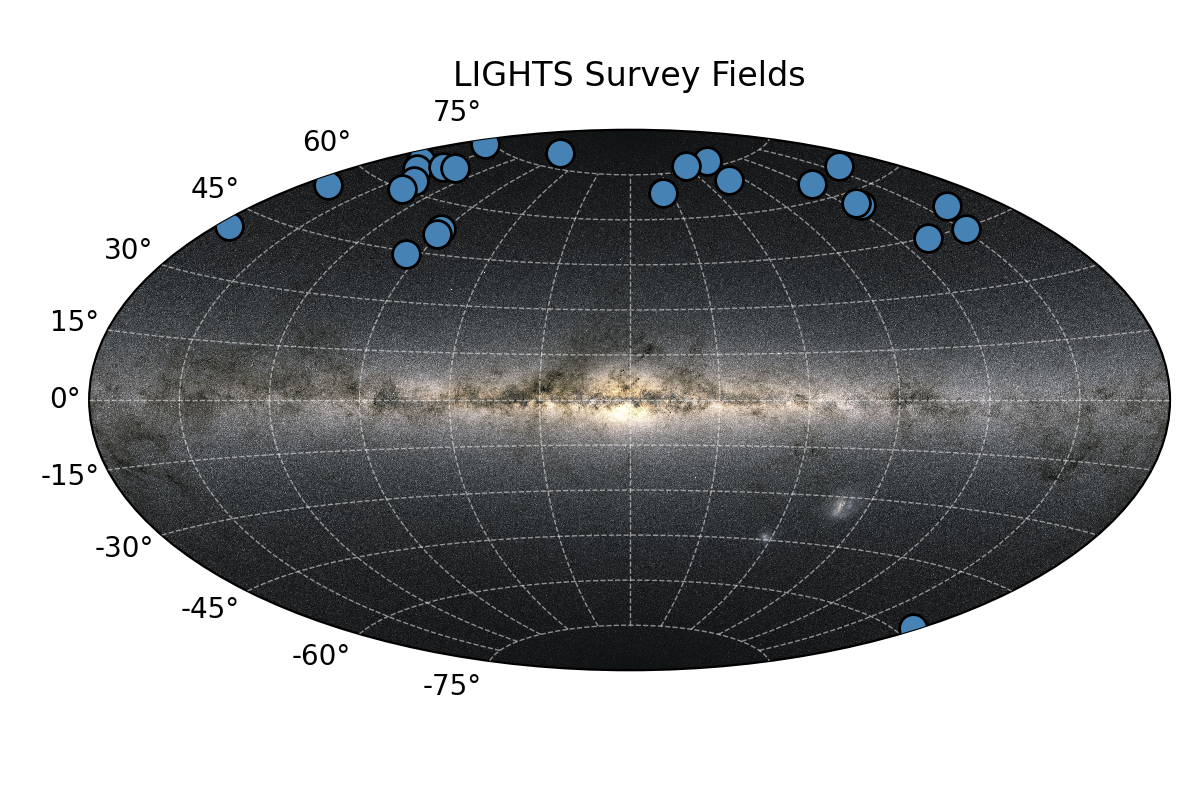}\\
\caption{The location of the 25 LIGHTS fields analyzed in this work in Galactic coordinates is shown here.}
    \label{fig:typicaldistatlimits_g_r}
\end{figure}

The spatial distribution of the analyzed LIGHTS fields in Galactic coordinates is presented in Fig. \ref{fig:typicaldistatlimits_g_r}. 

\section{Building the Hubble Ultra Deep Field  at ground-based spatial resolution}

\begin{figure*}
    \includegraphics[width=1\linewidth]{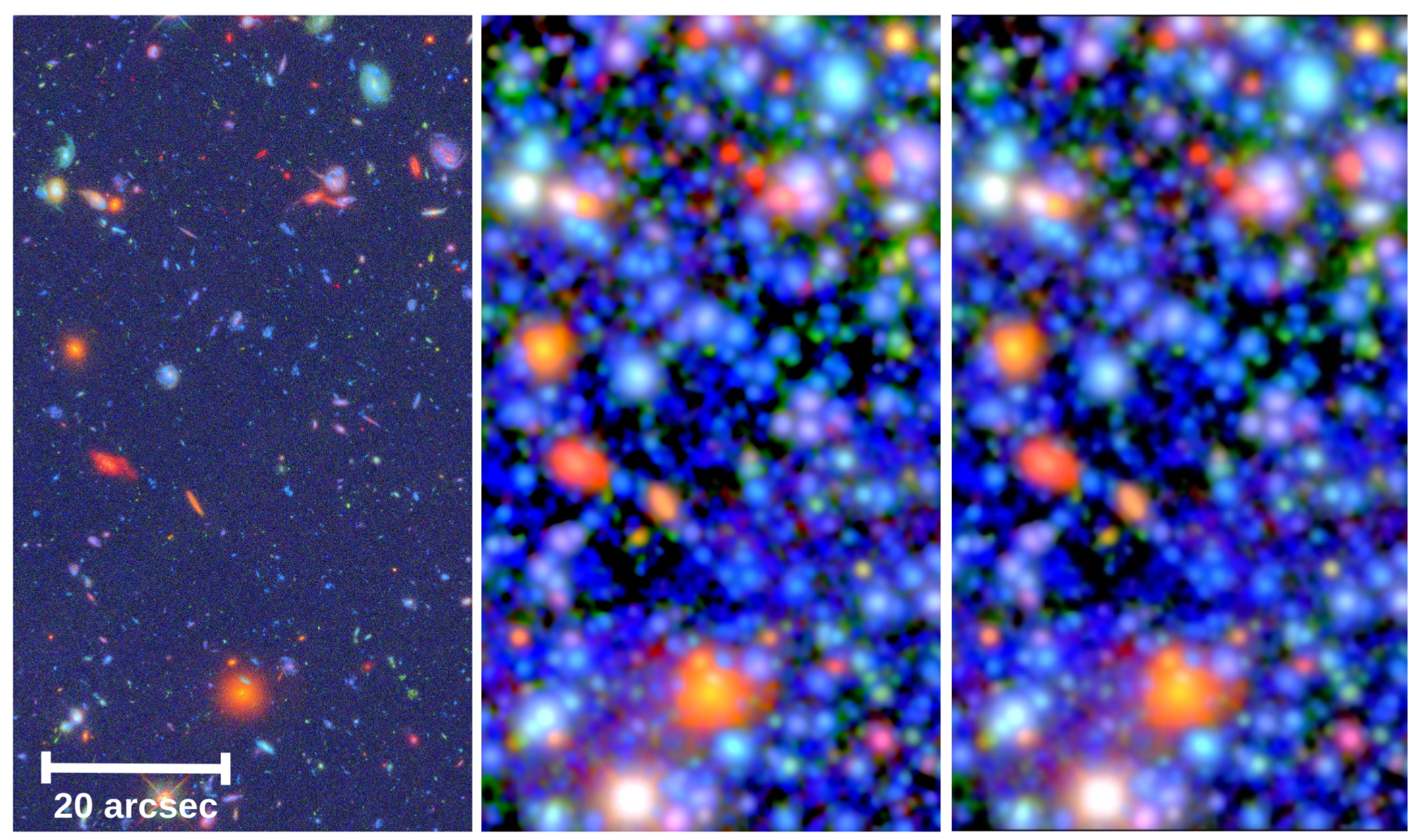}
    \caption{Color composite image (B(F435W), V(F606W), and z(F850LP) bands) of a section of the Hubble Ultra Deep Field image \citep{Beckwith2006}.  From left to right: The HUDF data at its original spatial resolution; In the middle panel, convolved with the LIGHTS PSF with the HUDF pixel scale of 0.03 arcsec. On the right is the image convolved and re-binned to the LIGHTS pixel scale, which has the same resolution as ground-based data.}
    \label{fig:typical_distance_HUDF}
\end{figure*}

To explore the fraction of pixels that the galaxies in the Hubble Ultra Deep Field will cover when observed at a spatial resolution from the ground, we proceed as follows: First, we downloaded the F435W, F606W, and F850LP images from \citet{Beckwith2006}. Then, we convolved the Hubble images with the PSF of LIGHTS \citep{Sedighi2025}, with a pixel scale of 0.03 from HUDF images. The result is shown in the middle panel of Fig. \ref{fig:typical_distance_HUDF}. The FWHM of the LIGHTS PSF is much larger than that of Hubble, so convolution of the HUDF images with the LIGHTS PSF basically reproduces ground-based conditions. Next, we resampled the HUDF images (while conserving the total flux) to the pixel scale of LIGHTS (i.e., 0.224 arcsec). This is shown in the right panel of Fig. \ref{fig:typical_distance_HUDF}. In this work, we performed quantitative estimations of the area occupied using F606W, the filter closest to the Sloan r-band. However, Fig. \ref{fig:typical_distance_HUDF} color-combines the three HST filters to produce an impression of the field's color at different resolutions.

 The Hubble Ultra Deep Field images were not processed using a low surface brightness-friendly pipeline. Therefore, they show background artifacts with patchy regions that are brighter in some bands and less visible in others. Additionally, the visual extension of the largest galaxies is severely underestimated due to oversubtraction, as demonstrated in \citet{2019A&A...621A.133B}.

\section{Expected Source Confusion in space-based observations}

\begin{figure}
    \centering
    \includegraphics[width=0.65\linewidth]{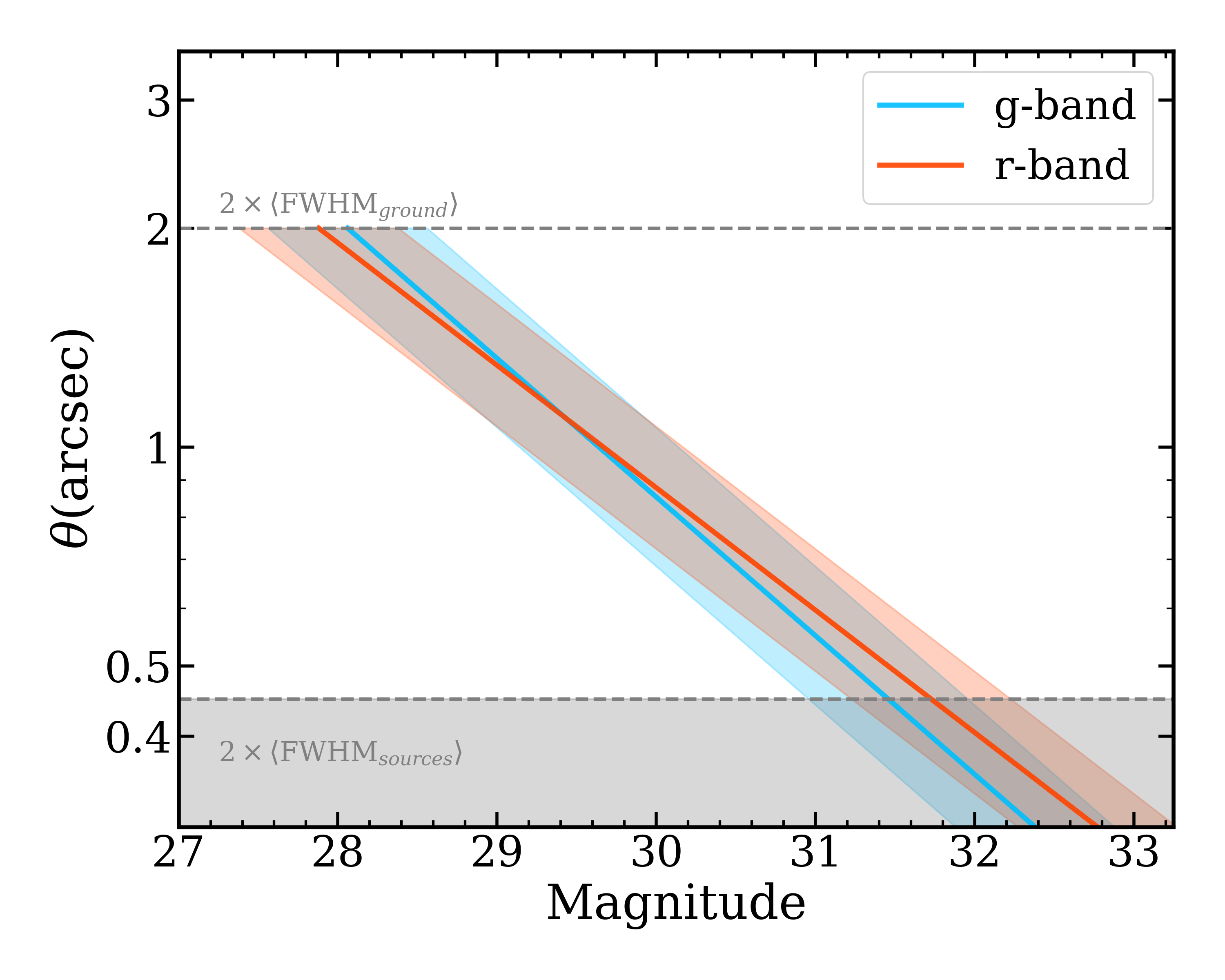}
    \caption{Extrapolated behavior of the typical angular distance between sources as the completeness magnitude of the images improves. In space and on the ground, aided by adaptive optics, the ultimate source confusion limit is not imposed by the spatial resolution of the images, but rather by the sources' intrinsic extension in the high-redshift universe. The typical extension of the very faint galaxies in the deepest Hubble data (as characterized by the FWHM) ranges from 0.15 to 0.225 arcsec. Therefore, the ultimate source confusion limit will eventually be reached when the typical separation is twice that size. This is expected to occur at a completeness magnitude between 31.5 to 32.6 mag. We have assumed a conservative uncertainty of $\pm$0.5 mag in the extrapolation values.}
    \label{fig:error_SCLM}
\end{figure}

The angular distance between sources, as measured using the Hubble deep catalogs, closely matches the expected extrapolation from the LIGHTS data collected from the ground (see right panel of Fig. \ref{fig:numbercountwithHST}). The natural question, therefore, is: At what point would source confusion occur if we had very deep images taken from space? As we have seen, the answer using ground-based data and natural seeing is dictated by seeing size. In space, however, the limiting factor is not the spatial resolution of the images, which could theoretically be improved with larger-aperture telescopes, but rather the apparent size of the natural sources. One characteristic of the current consensus cosmological model is that the apparent size of a given physical object does not decrease with redshift; rather, it remains fairly constant with z.
Therefore, to provide a rough estimate of the limiting magnitude at which we would expect source confusion from space, we proceeded as follows: First, we calculated the average apparent size of the sources in the Hubble Ultra Deep Field catalog. We characterized this size using the FWHM value provided in the catalog available in STScI\footnote{\url{https://archive.stsci.edu/pub/hlsp/udf/acs-wfc/}} for HUDF \citep{Beckwith2006}. Next, we estimated 2$\times$FWHM as the minimum distance at which two sources can be located for source confusion to occur. The measured FWHM value for the typical source in the HUDF (F606W) image is 0.225 arcsec. However, we note that other authors have found smaller sizes for very faint galaxies. For example, in the F606W band, \citet{2022ApJ...940L..15K} find a median size of 0.16 arcsec for their faintest objects in their Fig. 4. At slightly longer wavelengths (F090W) with JWST, \citet{2023AJ....165...13W} find a typical size of 0.15 arcseconds. When reaching such faint magnitudes, the size of the objects becomes comparable to the spatial resolution of our current space telescopes, which affects the size of faint objects. Fig. \ref{fig:error_SCLM} shows the range of measurements of the sizes of very faint galaxies obtained using space telescopes.

Fig. \ref{fig:error_SCLM} shows the extrapolation of the expected angular distance between sources based on the LIGHTS data as it extends to fainter and fainter magnitudes. According to this extrapolation, the ultimate limit for source confusion will be reached when any observation, whether from the ground using adaptive optics or from space, reaches a completeness magnitude between 31.5 to 32.6 mag. This completeness magnitude exceeds our current observational capabilities, but larger aperture telescopes\footnote{Very deep observations using JWST obtain a 5$\sigma$ limiting magnitude for point-like sources (0.2 arcsec apertures) of 30.85 mag (F090W) after 39.1 hours \citep{2025arXiv251107542A}. Assuming that the completeness magnitude for both point-like and extended sources is more than 1 mag brighter than the  5$\sigma$ limiting magnitude for point-like sources,  current technology is still far from the ultimate source confusion limit.} in the space during this century could eventually reach it.

%--------------------------------------------------------------------------
\end{document}